\documentclass{article}

\usepackage{amsthm}
\usepackage{newtxtext,newtxmath}

\usepackage{graphicx}
\usepackage{subcaption}
\usepackage[table]{xcolor}
\usepackage{mdframed}
\usepackage{booktabs}
\usepackage{multirow}
\usepackage{tabularx}
\newcolumntype{C}[1]{>{\hsize=#1\hsize}X}
\usepackage{url}
\usepackage{bm}
\usepackage{gensymb}

\newcommand{\transpose}{\mkern-1.5mu\mathsf{T}}
\newcommand{\quotesep}{\vspace{-0.5em}\noindent\hfil\rule[0.5ex]{0.25\textwidth}{0.4pt}\vspace{-0.5em}}

\begin{document}

\title{A qualitative field study on explainable AI for lay users subjected to AI cyberattacks}

\author{
    Kevin McAreavey$^1$ \and Weiru Liu$^1$ \and Kim Bauters$^1$ \and
	Dennis Ivory$^2$ \and George Loukas$^2$ \and Manos Panaousis$^2$ \and Hsueh-Ju Chen$^2$ \and
	Rea Gill$^3$ \and Rachael Payler$^3$ \and 
	Asimina Vasalou$^4$
}

\date{\small
    $^1$University of Bristol, Bristol, UK\\
	\{kevin.mcareavey, weiru.liu, kim.bauters\}@bristol.ac.uk\\
	\vspace{1em}
    $^2$University of Greenwich, London, UK\\
	\{dennis.ivory, g.loukas, e.panaousis, hsueh-ju.chen\}@greenwich.ac.uk\\
	\vspace{1em}
	$^3$University of Reading, Reading, UK\\
	\{r.k.gill, r.l.payler\}@student.reading.ac.uk\\
	\vspace{1em}
	$^4$University College London, London, UK\\
	a.vasalou@ucl.ac.uk
}

\maketitle

\begin{abstract}
	In this paper we present results from a qualitative field study on explainable AI (XAI) for lay users ($n = 18$) who were subjected to AI cyberattacks.
	The study was based on a custom-built smart heating application called Squid and was conducted over seven weeks in early 2023.
	Squid combined a smart radiator valve installed in participant homes with a web application that implemented an AI feature known as setpoint learning, which is commonly available in consumer smart thermostats.
	Development of Squid followed the XAI principle of interpretability-by-design where the AI feature was implemented using a simple glass-box machine learning model with the model subsequently exposed to users via the web interface (e.g.\ as interactive visualisations).
	AI attacks on users were simulated by injecting malicious training data and by manipulating data used for model predictions.
	Research data consisted of semi-structured interviews, researcher field notes, participant diaries, and application logs.
	In our analysis we reflect on the impact of XAI on user satisfaction and user comprehension as well as its use as a tool for diagnosing AI attacks.
	Our results show only limited engagement with XAI features and suggest that, for Squid users, common assumptions found in the XAI literature were not aligned to reality.
	On the positive side, users appear to have developed better mental models of the AI feature compared to previous work, and there is evidence that users did make some use of XAI as a diagnostic tool.
\end{abstract}

\section{Introduction}\label{sec:introduction}
Neglect of user studies has been described as the ``original sin'' of explainable AI (XAI) research~\cite{Keane:IJCAI:2021} with the XAI community itself regularly advocating the need for more user studies~\cite{Keane:IJCAI:2021,Miller:XAI:2017,VanDerWaa:AIJ:2021,Verma:ML-RSA:2020}.
An obvious challenge in addressing this deficit is that user studies are costly, both in time and resources~\cite{Doshi-Velez:arXiv:2017,Miller:XAI:2017}.
Another challenge is that AI systems tend to exhibit a wide range of stakeholders yet the needs of different classes of stakeholder may be very different~\cite{Arrieta:IF:2020,Langer:AIJ:2021}.
For example, lay users are an important but under-represented class of stakeholder in XAI~\cite{Zylowski:XI-ML:2022}; they can be understood as end-users with no particular expertise in AI who are nonetheless affected by AI decisions.
A particular challenge with lay users is that basic assumptions around numeracy, literacy, and technological competence cannot be taken for granted~\cite{Cheng:CHI:2019,Correll:CHI:2019}.
This suggests that the appropriateness of certain kinds of user studies may be very different for lay users compared to other stakeholders.
At the same time, lay users may be less able than other stakeholders to protect themselves against the negative effects of AI~\cite{Yang:UbiComp:2013}, so developing appropriate XAI for lay users seems crucial to the broader aims of ethical AI.

In this paper we present results from a qualitative field study on XAI for lay users conducted over seven weeks in early 2023.
The study involved 18 participants from 10 households in London and south-west England.
To enable the study, we developed a consumer-grade smart heating application, called Squid, that combined consumer hardware with a custom-built web application.
The hardware was a thermostatic radiator valve (TRV), which is a device that controls the flow of hot water into a radiator so as to maintain room temperature relative to a user-specified target, called a setpoint.
The web application included hardware integration and a web interface but also replicated an AI feature found in consumer smart thermostats, known as setpoint learning.
Commercial examples of setpoint learning include the Auto-Schedule feature from the Nest Learning Thermostat and the eco+ feature from ecobee.
The aim of setpoint learning is to adjust setpoints automatically by learning user preferences so as to reduce the need for manual interventions while maintaining user satisfaction.
In replicating this feature our intention was to focus the study on AI functionality that lay users might encounter in the real-world.
However, the AI technology used by commercial implementations of setpoint learning remains a trade secret~\cite{Barrett:ECML-PKDD:2015}, so the design of Squid was instead based on academic work previously published in leading AI venues~\cite{Alan:CHI:2016,Shann:AAMAS:2017,Shann:IJCAI:2013,Shann:AAMAS:2014}.
Unlike most commercial examples of setpoint learning, the feature in Squid was designed for households that pay for energy according to dynamic energy tariffs, such as those offered by Octopus Energy in the UK.
Under these tariffs, prices vary in 30-minute slots in line with wholesale prices and thus allow consumers to reduce energy costs by shifting energy consumption from when prices are high to when prices are low.
The AI feature in Squid exploits this cost-saving potential by automating frequent setpoint adjustments in response to price changes, adjustments that would otherwise be impractical under manual control.
The original idea was said to ``trade off comfort versus cost''~\cite{Shann:AAMAS:2017}.

Development of Squid can be understood as having followed the XAI principle of interpretability-by-design~\cite{Rudin:NMI:2019} where the AI feature was implemented using a glass-box (or inherently interpretable) machine learning model with the model exposed to users via the web interface in the form of e.g.\ interactive visualisations~\cite{Yang:HCIS:2023}.
A novel angle of the field study was a focus on the nascent cyberattack vector of AI attacks~\cite{Pitropakis:CSR:2019,Szegedy:ICLR:2014}: during the study participants were subjected to three kinds of simulated AI attacks, all intended to affect the user via the setpoint learning feature.
Attacks included the injection of malicious training data (poisoning attacks) and the manipulation of energy prices used by the models to make predictions (evasion attacks).
Poisoning attacks mimicked situations where an attacker gains access to the system and injects data using standard controls (simple attacks) but also those where an attacker takes measures to conceal their activity (complex attacks).
Research data consisted of semi-structured interviews, researcher field notes, participant diaries, and application logs.
In our analysis we reflect on the impact of XAI on user satisfaction and user comprehension as well as its use as a tool for diagnosing AI attacks.
Our results show only limited engagement with XAI features and suggest that, for Squid users, common assumptions found in the XAI literature were not aligned to reality.
On the positive side, users appear to have developed better mental models of the AI feature compared to previous work, and there is evidence that users did make some use of XAI as a diagnostic tool.

The rest of the paper is organised as follows:
in Section~\ref{sec:related_work} we review existing user studies from the XAI literature;
in Section~\ref{sec:methodology} we present our methodology, including the design of Squid and the field study itself;
in Section~\ref{sec:results} we present our results;
in Section~\ref{sec:discussion} we reflect on limitations and lessons learned;
and in Section~\ref{sec:conclusion} we conclude.
For preliminary work on the development of Squid see~\cite{McAreavey:ICAART:2022}.
For a separate strand of research associated with the same field study see~\cite{Vasalou:IJHCS:2024}.

\section{Related Work}\label{sec:related_work}
In a survey of the literature on XAI evaluations, Nauta et al.~\cite{Nauta:CS:2023} identified 312 papers published between 2014--2020 and found that only 22\% had reported user studies.
In a survey specifically on XAI evaluations with users, Rong et al.~\cite{Rong:PAMI:2024} identified a total of 97 user studies published between 2018--2022.
The authors conducted a meta-analysis of the results reported by these user studies and identified the following trends:
(i) XAI is effective at increasing \emph{subjective} comprehension where a user believes that their mental model is correct;
(ii) it remains unclear whether XAI is effective at increasing user trust and/or satisfaction;
(iii) XAI is not effective at convincing users that an AI system is fair; and
(iv) interactiveness in XAI has a positive effect on user trust, satisfaction, and comprehension.
On reflection, the authors point out that merely increasing subjective comprehension is not necessarily desirable, since humans are known to overestimate their understanding of complex systems~\cite{Rozenblit:CS:2002}.
The authors also raise an issue that may be of particular relevance to our user study, especially in the context of AI attacks: ``good explanations [...] reveal weaknesses of the model'' and ``users may express their negative feelings about the model through negative ratings of the explanations''.
The authors did not differentiate qualitative and quantitative user studies but supplementary material\footnote{\url{https://doi.org/10.1109/TPAMI.2023.3331846/mm1}} identifies three studies as making use of semi-structured interviews~\cite{Ehsan:CHI:2021,LeBras:CHI:2018,Li:IUI:2019}.

In a survey of sample sizes in XAI user evaluations, Peters et al.~\cite{Peters:IEEE:2023} identified a total of 220 user studies published between 2012--2022.
The authors were specifically interested in whether reported results were supported by sample size and found that over-generalisations were pervasive: most studies did not justify sample size, most generalised results beyond samples, and none provided evidence that findings correlated with larger samples.
Although qualitative studies do not typically aim for generalisability~\cite{Polit:IJNS:2010}, the authors argue that sample size justification is still important for qualitative studies, e.g.\ to ensure saturation~\cite{Vasileiou:MRM:2018}.
In total, the survey identified 126 quantitative studies (57\%) and 27 qualitative studies (12\%) with the rest being a mixture of the two.
Supplementary material\footnote{\url{https://osf.io/vzndw/}} shows that six qualitative studies (22\%) targeted lay users~\cite{Guesmi:IUI-WS:2022,Jin:arXiv:2022,Mai:IUI:2020,Schroder:Perspectives:2021,Szymanski:IUI-WS:2022,Zhang:MobiTAS:2022} and a further two studies (7\%) targeted a mixture of lay users and technology experts~\cite{Dieber:IF:2022,Spinner:VCG:2020} with the rest targeting technology experts and/or domain experts.
Qualitative studies targeting lay users were thus very much in the minority.
Median sample size for these eight studies was $n = 11$ and all sample sizes were in the range $9 \le n \le 13$ except for one outlier~\cite{Jin:arXiv:2022} with $n = 32$.
The picture is very different for quantitative studies where 110 targeted lay users (87\%) and a further 7 targeted a mixture of lay users and technology experts (6\%) with only 6 targeting technology experts and/or domain experts (5\%).
These differences are perhaps unsurprising given that lay users represent a larger population than experts, yet lay users may be less willing than experts to commit to the high demands typically associated with qualitative studies.
Supplementary material suggests that mixed studies shared more in common with quantitative studies than with qualitative studies, including samples sizes up to $n = \text{1,118}$ and a majority targeting lay users; these studies might thus be better understood as quantitative studies with a qualitative component rather than vice versa.

\section{Methodology}\label{sec:methodology}
In this section we describe our methodology, including the design of Squid (AI feature, XAI features, hardware, software, user interface) and the design of the field study itself (simulated AI attacks, recruitment, logistics, data collection).

\subsection{AI Feature}\label{sec:ai}
In the AI literature the only example of setpoint learning that has been validated with end-users appears to be a series of work by Shann et al.~\cite{Alan:CHI:2016,Shann:AAMAS:2017,Shann:IJCAI:2013,Shann:AAMAS:2014}.
Of particular relevance in their work is a field study conducted with lay users focusing on custom-built smart heating application called SmartThermo~\cite{Alan:CHI:2016,Shann:AAMAS:2017}.
The setpoint learning feature in SmartThermo uses Bayesian linear regression~\cite[Chapter 3]{Bishop:book:2006} such that user preferences are learned as a mapping from (dynamic) energy prices to setpoints.
The design is as follows.
When a user manually adjusts the setpoint, the current energy price and chosen setpoint are taken together as input and used to update the Bayesian model.
While the model is active, the setpoint is automatically adjusted by the system according to the model's prediction for the current energy price.
Three alternative end-user designs were then considered during the field study with the results identifying one of those designs, called indirect learning, as being most-preferred by users.
According to this design each input immediately updates the model but the model is then temporarily overridden for a period of e.g.\ 30 or 60 minutes.
During this override period the original setpoint specified by the user remains active.
When the override period ends, the (updated) model becomes active once again.
In an alternative design called direct learning, user setpoint adjustments only served to update the model and were otherwise ignored by the system, but this design was found to frustrate users.
Squid thus replicates the indirect learning design, as well as the use of Bayesian linear regression.

We will now outline the setpoint learning method used in Squid.
For further details see~\cite{Alan:CHI:2016,Shann:AAMAS:2017}.
Let $m_{i} \sim \mathcal{N}(\pmb{\mu_{i}}, \pmb{\Sigma_{i}})$ be a (conjugate) prior distribution where $\pmb{\mu_{i}}$ is the mean and $\pmb{\Sigma_{i}}$ is the covariance matrix.
An input is a pair $(\pmb{x}, y)$ where $\pmb{x} \in \mathbb{R}^{n}$ is a datapoint and $y \in \mathbb{R}$ is a label.
The posterior distribution $m_{i+1} \sim \mathcal{N}(\pmb{\mu_{i+1}}, \pmb{\Sigma_{i+1}})$ is defined as:
\begin{align}
	\pmb{\Sigma_{i+1}^{-1}} & = \pmb{\Sigma_{i}^{-1}} + \beta \pmb{x_{i}}^{\transpose} \pmb{x_{i}} \\
	\pmb{\mu_{i+1}} & = \pmb{\Sigma_{i+1}} \left( \pmb{\Sigma_{i}^{-1}} \pmb{\mu_{i}} + \beta \pmb{x_{i}}^{\transpose} y_{i} \right)
\end{align}
where $\pmb{\Sigma_{i}^{-1}}$ is the inverse covariance matrix and $\beta \in \mathbb{R}$ is the input noise precision.
The predictive distribution $y_{i}(\pmb{x}) \sim \mathcal{N}(\mu_{y_{i}}, \sigma_{y_{i}}^{2})$ for each datapoint $\pmb{x}$ is defined as:
\begin{align}
	\mu_{y_{i}} & = \pmb{\mu_{i}}^{\transpose} \pmb{x} \\
	\sigma_{y_{i}}^{2} & = \frac{1}{\beta} + \pmb{x}^{\transpose} \pmb{\Sigma_{i}} \pmb{x} \
\end{align}
The initial distribution in Squid, also known as the default model, is a bivariate distribution $m_{1} \sim \mathcal{N}(\pmb{\mu_{1}}, \pmb{\Sigma_{1}})$ defined as:
\begin{align}
	\pmb{\mu_{1}} & = \begin{bmatrix}
		\mu_{b_{1}} \\
		\mu_{s_{1}}
	\end{bmatrix} \\
	\pmb{\Sigma_{1}} & = \begin{bmatrix}
		\sigma_{b_{1}}^{2} & \rho \sigma_{b_{1}} \sigma_{s_{1}} \\
		\rho \sigma_{b_{1}} \sigma_{s_{1}} & \sigma_{s_{1}}^{2}
	\end{bmatrix}
\end{align}
where $\mu_{b_{1}} \in \mathbb{R}$ is the initial bias (mean), $\mu_{s_{1}} \in \mathbb{R}$ is the initial slope (mean), and $\rho \in \mathbb{R}$ is the correlation coefficient.
Each input $(\pmb{x_{i}}, y_{i})$ serves to update the prior distribution $m_{i}$ giving the posterior distribution $m_{i+1}$ with $m_{i+1}$ then becoming the prior for the next update; this is an example of online learning~\cite{Hoi:NC:2021}.
In Squid the label $y_{i}$ represents a new setpoint chosen by the user and $\pmb{x_{i}}$ represents the energy price when the input was made.
The following hyperparameters were chosen based on observed learning dynamics under typical usage: $\mu_{b_{1}} = 22$, $\mu_{s_{1}} = -0.245$, $\sigma_{b_{1}}^{2} = 1$, $\sigma_{s_{1}}^{2} = 0.01$, $\rho = -0.3$, and $\beta = 0.33$.
Similar to~\cite{Alan:CHI:2016,Shann:AAMAS:2017} we refer to $\mu_{b_{1}}$ as the \textbf{preferred temperature (if energy were free)} and to $\mu_{s_{1}}$ as the \textbf{price sensitivity}.
A model's setpoint predictions for a time series of energy prices is called a setpoint schedule.

\subsection{XAI Features}\label{sec:xai}
A machine learning model is a glass-box model (or inherently interpretable model) if its internals can be inspected and ascribed meaning that is intuitive to humans (e.g.\ decision trees, linear models).
Conversely, a machine learning model is a black-box model if its internals cannot be inspected or can be inspected but cannot be ascribed intuitive meaning (e.g.\ neural networks, ensemble models).
XAI for machine learning is commonly divided between interpretability-by-design (i.e.\ use glass-box models) and post-hoc methods (i.e.\ explain/interpret any already-trained model)~\cite{Abdul:CHI:2018,Rudin:NMI:2019}.
It is commonly accepted that there is a trade-off between interpretability and performance, with black-box models being more performant than glass-box models~\cite{Das:arXiv:2020}.
Choosing interpretability-by-design can thus be understood as sacrificing performance for the sake of interpretability.
However, the fact that a model may be inherently interpretable does not imply that it is inherently \emph{comprehensible}, although these two concepts are often conflated in the literature~\cite{Lipton:ACMQ:2018}.
A classic example is a decision tree with thousands of nodes: it is a glass-box model consisting of a tree where nodes have intuitive meaning (e.g.\ if $X > Y$ then go to child $A$ else go to child $B$) and a path can be easily traced to a prediction, yet humans are unlikely to comprehend a tree of that size.
Likewise, the ability to inspect a glass-box model does not explain \emph{why} the model looks as it does, so interpretability-by-design in itself does not avoid other challenges facing XAI.

As a linear model, Bayesian linear regression would typically be regarded as a glass-box model~\cite{Rudin:NMI:2019}.
Squid can thus be said to follow the approach of interpretability-by-design, albeit as a consequence of SmartThermo having implicitly followed the same approach.
Squid relies on a single input feature (price) and a single target feature (setpoint), which gives rise to only two model parameters (bias and slope).
This places Squid arguably among the simplest instances of a particular kind of glass-box model.
In turn this poses an interesting research question: can lay users understand what may be regarded by AI experts as an especially simple glass-box model?
If not, it seems unlikely that lay users would fair much better with a more performant black-box model, irrespective of the choice of post-hoc method(s).
As mentioned, interpretability-by-design does not itself avoid other challenges facing XAI, and this is especially true in the context of lay users.
Two major challenges in Squid are: (i) that the model expresses (probabilistic) uncertainty over learned user preferences and (ii) that the model evolves over time based on user inputs, a predefined prior distribution $m_{1}$, and an input noise parameter $\beta$.
How to communicate this information to lay users---in a way that is accurate yet relevant to their needs---is far from clear, given that it relies on mathematical concepts that may be unfamiliar to them.

Popular post-hoc methods in XAI include: 
global feature dependency methods~\cite{Friedman:AS:2001,Apley:JRSSS:2020}, which measure the relationship between input features and model predictions; 
local feature importance methods~\cite{Lundberg:NIPS:2017,Ribeiro:KDD:2016}, which measure the importance of input features on the prediction for a given datapoint; 
and counterfactual example methods~\cite{Mothilal:FAT:2020,Wachter:JOLT:2018}, which select alternative datapoints leading to some desired prediction.
All three methods would trivialise in Squid, e.g.\ feature-based methods would reduce to a single input feature (price) while counterfactual example methods would reduce to inspecting a two-dimensional line chart (price vs.\ setpoint) that is also linear.
An obvious alternative then is to expose the glass-box model to users with the aid of visualisations~\cite{Gunning:AIM:2019,Biran:XAI:2017}, especially interactive visualisations~\cite{Abdul:CHI:2018}.
This is the approach taken in Squid, which is supported by previous findings on the benefits of interactiveness in XAI (see Section~\ref{sec:related_work}).

\begin{figure}[p]
	\centering
	\includegraphics[width=0.5\textwidth]{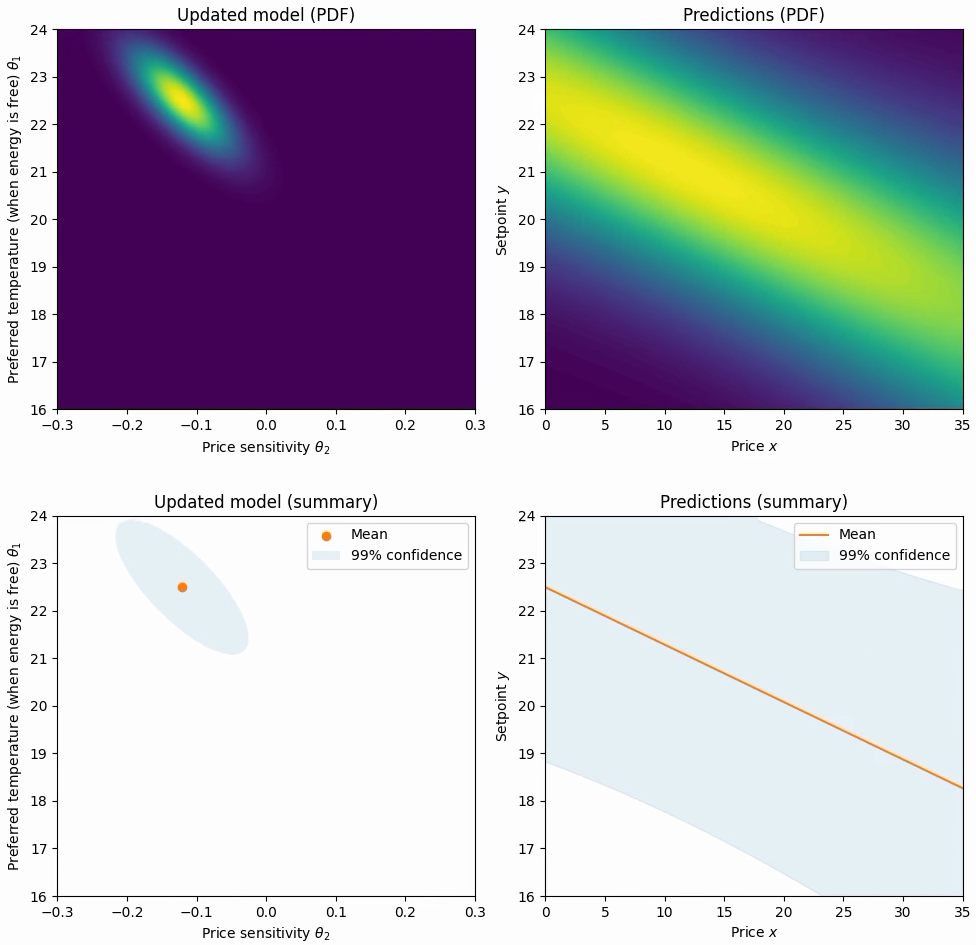}
	\caption{Example visualisations of model parameters and model predictions, including (top) heatmaps and (bottom) summary statistics.}
	\label{fig:heatmap}
\end{figure}

\begin{figure}[p]
	\centering
	\begin{subfigure}{\textwidth}
		\centering
		\includegraphics[width=\textwidth]{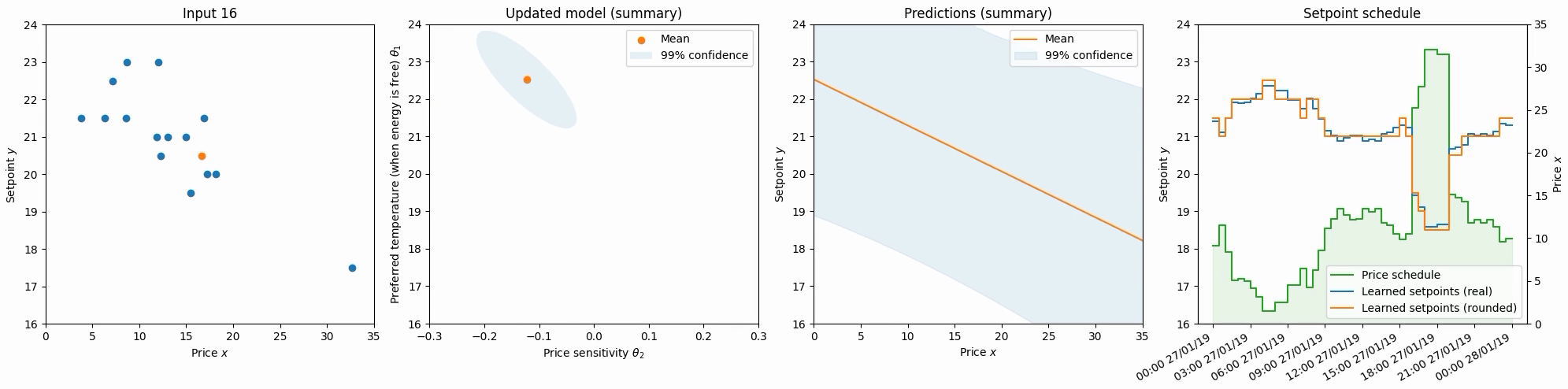}
		\caption{After 16 inputs.}
	\end{subfigure}%
	\vspace{1em}
	\begin{subfigure}{\textwidth}
		\centering
		\includegraphics[width=\textwidth]{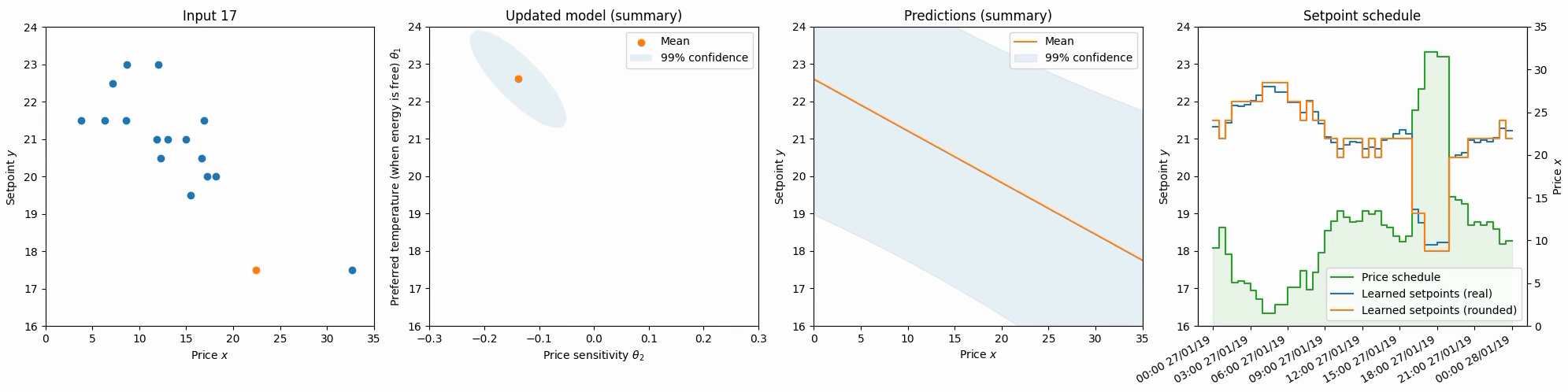}
		\caption{After 17 inputs.}
	\end{subfigure}
	\caption{Example visualisation of model evolution, including (left-to-right) user inputs, model parameters, model predictions, and a learned setpoint schedule.}
	\label{fig:interactive}
\end{figure}

Visualisations in Squid can be offered at various levels, including on input/training data, model predictions, and model parameters.
Such visualisations are well-established in data science and data analytics.
For example, Figure~\ref{fig:heatmap} visualises an example model and its corresponding predictions as outlined in Section~\ref{sec:ai}, with the top row using heatmaps to visualise probability densities and the bottom row using summary statistics (mean and confidence region) to visualise the same information.
More challenging is to convey how a model evolves over time in response to user inputs and its impact on the setpoint that would be adopted by Squid for some given price data.
Figure~\ref{fig:interactive} illustrates how a set of interactive charts might convey such information by allowing the user to step through their history of inputs and to inspect its effect on model parameters, model predictions, and the learned setpoint schedule.
The XAI features in Squid include the kinds of charts outlined in Figures~\ref{fig:heatmap}--\ref{fig:interactive} with the addition of (i) notification logs that use natural langauge to convey changes to the model and (ii) a graphic known as the \emph{gauge} that separately visualises the price sensitivity parameter.
Further details on these features will be elaborated in Section~\ref{sec:software} when presenting the user interface.

\subsection{Hardware}\label{sec:hardware}

\begin{figure}[tbp]
	\centering
	\includegraphics[width=0.45\textwidth]{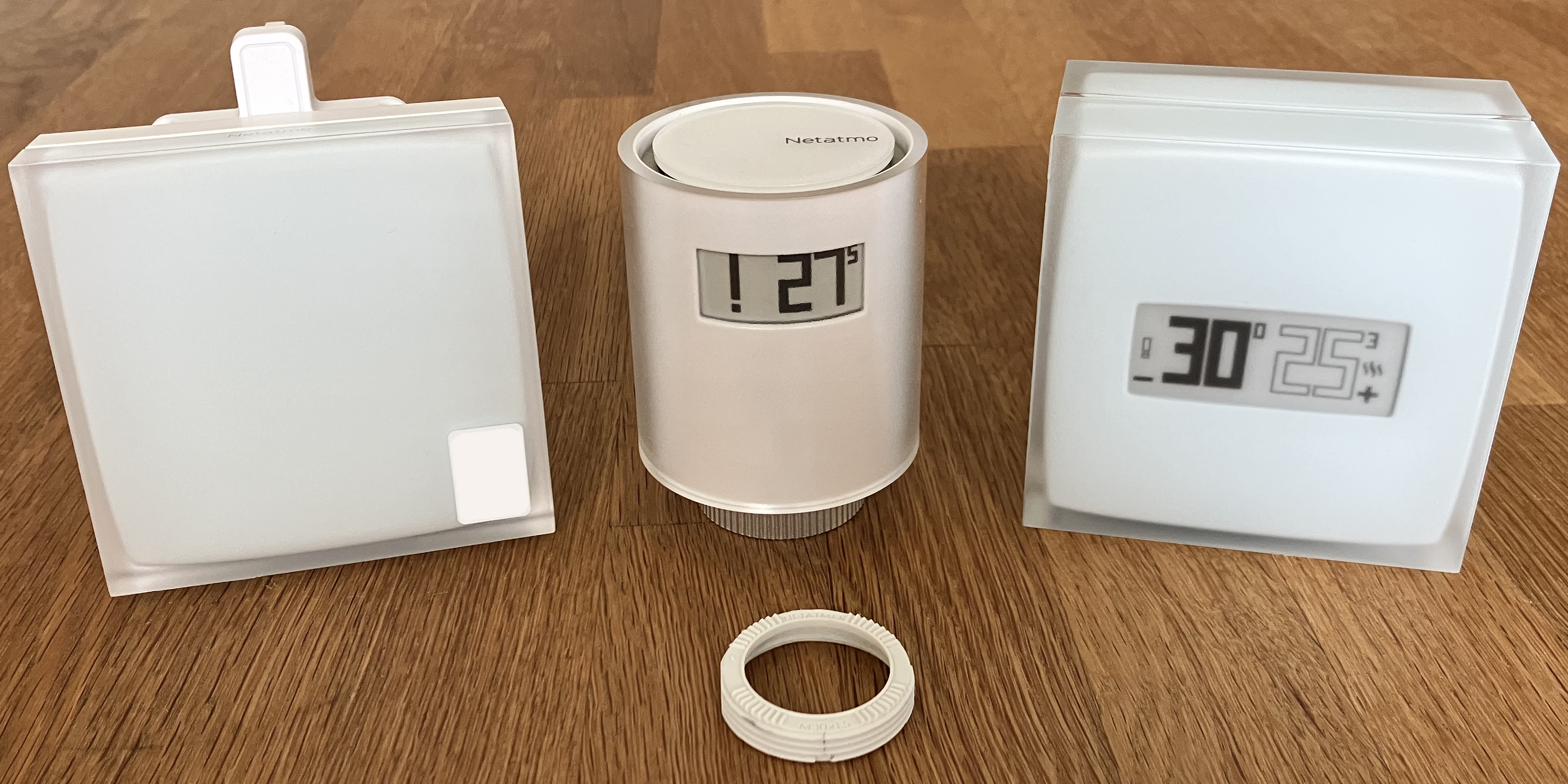}
	\caption{Netatmo hardware, including (left-to-right) relay with mains adapter, smart radiator valve with (below) radiator adapter, and smart thermostat.}
	\label{fig:hardware}
\end{figure}

Several consumer smart TRVs were investigated, including the Lightwave LW922, the Meross MTS150H, and the Netatmo Additional Smart Radiator Valve.
Ultimately the Netatmo device was chosen due to its reliability, availability of stock, and strength of technical support.
We often refer to this device as the \textbf{valve}.
Netatmo provide simple instructions on how to attach the valve to a radiator using a bundled radiator adapter, replacing any TRV already attached. 
Netatmo also provide a public API giving remote hardware access, including access to on-board temperature readings and the ability to adjust setpoints.
Squid was responsible for choosing setpoints but the actual control mechanism used to maintain room temperature relative to those setpoints was otherwise delegated to Netatmo.
Although Squid was designed for this particular TRV, in principle it could be easily adapted to work with any consumer smart TRV or smart thermostat.
The valve communicates with Netatmo servers via a bundled wireless hub, which is typically wired to a boiler but can also connect to a mains socket using a bundled mains adapters.
We refer to this device as the \textbf{relay}; during the field study it was connected to the mains.
In order for the valve to operate correctly, Netatmo expects that another device called the Netatmo Smart Thermostat is also present and connected to the relay.
Typically this device functions as a standard thermostat (i.e.\ it acts as boiler controller) but in our setup the device only functioned as an additional temperature sensor, with its main purpose being to ensure the valve operated correctly.
We refer to this device as the \textbf{cube}.
The full collection of Netatmo hardware is shown in Figure~\ref{fig:hardware}, which comprises the set of hardware installed in each participant home.
The valve and cube include on-board displays showing their current temperature reading and setpoint.
However, since the cube was used neither as an actuator nor as a primary sensor, a sticker was placed over its display prior to installation.
Participants were told that the cube should remain in the same room as the valve but that it could otherwise be ignored.
Finally, each household was given a standard Galaxy Tab A7 Lite (8.7 inch) tablet for accessing the Squid web application.
This tablet has a scaled resolution of 893 $\times$ 533 pixels.

\subsection{Software \& User Interface}\label{sec:software}
The software architecture for Squid consists of five components (see below).
All source code is available online under MIT license.\footnote{\url{https://github.com/chai-project}}
Components were deployed during the study on an Ubuntu 22.04 virtual machine hosted on a virtual private server located in the UK.
Standard security procedures were in place, including the isolation of deployed components among several Linux containers.
System and user data was stored in a PostgreSQL database with backups scheduled using Cron to run every day.
Energy price data was taken from the Agile tariff by Octopus Energy for 2019 in the London region, with dates offset to 2023, i.e.\ January 2019 data was used for January 2023, and so on.
The Agile tariff is capped at 35p/kWh and permits negatives prices, meaning that consumers might earn money to use energy.
However, all prices during our study were non-negative (min: 1.4, max: 35, mean: 12.5, SD: 5.9).
Live prices were not used due to the effect of the global energy crisis on dynamic tariffs during early 2023: for the Agile tariff, prices rarely fell below the 35p/kWh cap,\footnote{\url{https://octopus.energy/blog/the-state-of-wholesale-energy/}} which was an unprecedented situation that would have negatively affected the setpoint learning feature in Squid.
The five components of the software architecture are as follows:
\begin{itemize}
	\item \textbf{chai-frontend} is the user interface, which is accessed via a web browser.
	It is implemented in TypeScript using React and various libraries, including Chart.js for visualisations.
	During the study it was served using an Nginx web server.
	\item \textbf{chai-backend-api} is a web API that integrates with the Squid database and provides all backend functionality required by the user interface.
	It is implemented in Python using various libraries, including the Falcon web framework.
	During the study it was served through an Nginx reverse proxy.
	\item \textbf{chai-data-sources} is a custom Python wrapper for the Netatmo web API, which enables convenient access to hardware from other components.
	It is implemented in Python using various libraries.
	\item \textbf{chai-persistance} is a system service that monitors hardware (e.g.\ to capture temperature readings).
	It is implemented in Python using various libraries.
	During the study it was scheduled using Cron to run every five minutes.
	\item \textbf{chai-ai} is a system service that monitors the Squid database for user inputs, updates AI models, and (pre)generates associated XAI data.
	It is implemented in Python using various libraries, including SciPy to calculate confidence regions.
	During the study it was scheduled using Cron to run every minute.
\end{itemize}

The SmartThermo application by Shann et al.~\cite{Alan:CHI:2016,Shann:AAMAS:2017} included two features that allowed users to retain a degree of control over its AI feature.
The first was a \emph{boost} option, which allowed users to disable the AI feature temporarily, causing the system to enter an always-on mode.
The second was a \emph{settings page}, which in effect allowed users to reset their learned preferences, replacing their existing Bayesian model with a new prior distribution.
There is evidence that users made use of both features when they were unsatisified with their learned preferences, and that they did so on a frequent basis.
In the case of the direct learning design, for example, it was reported that users made on average less that three setpoint adjustments before opting to reset their learned preferences~\cite{Alan:CHI:2016}.
This suggests that users were frequently unsatisified with the learned models.
A possible explanation is that the design of SmartThermo failed to capture any temporality in user preferences: each user had a single Bayesian model that was active regardless of time-of-day or day-of-week, and the model itself did not consider any data that might give temporal context to user inputs.
In~\cite{Alan:CHI:2016} the authors acknowledge that input data was limited and imply that a lack of temporality in learned preferences was in fact raised by several participants, as seen in the following quote:
\begin{quote}
	But like I said, at night, I didn't want it so warm, though perhaps I quite sort of would like it to keep it a degree or two cooler when the temperature's high to save money or something like that.
\end{quote}
Arguably this issue is analogous to the distinction between a traditional thermostat and a \emph{programmable} thermostat, where the former relies on a setpoint that is fixed regardless of time-of-day or day-of-week, and the latter relies on a setpoint that follows a programmable setpoint \emph{schedule}, typically a rolling one-week schedule.
Following this idea, a reasonable generalisation of SmartThermo that is consistent with the design of programmable thermostats is to give users multiple Bayesian models and allow them to manually schedule when those models are active.
This design would capture some temporality in user preferences, including the temporal context of user inputs, without changing the underlying Bayesian method proposed by Shann et al.
We expect that this would improve user satisfaction in learned preferences over the design used by SmartThermo, so this is the design we adopt in Squid.
We will now outline the four main elements of the user interface, which variously reflect this design.

\begin{figure}[ptb]
	\begin{subfigure}{1.0\textwidth}
		\centering
		\includegraphics[width=0.45\textwidth]{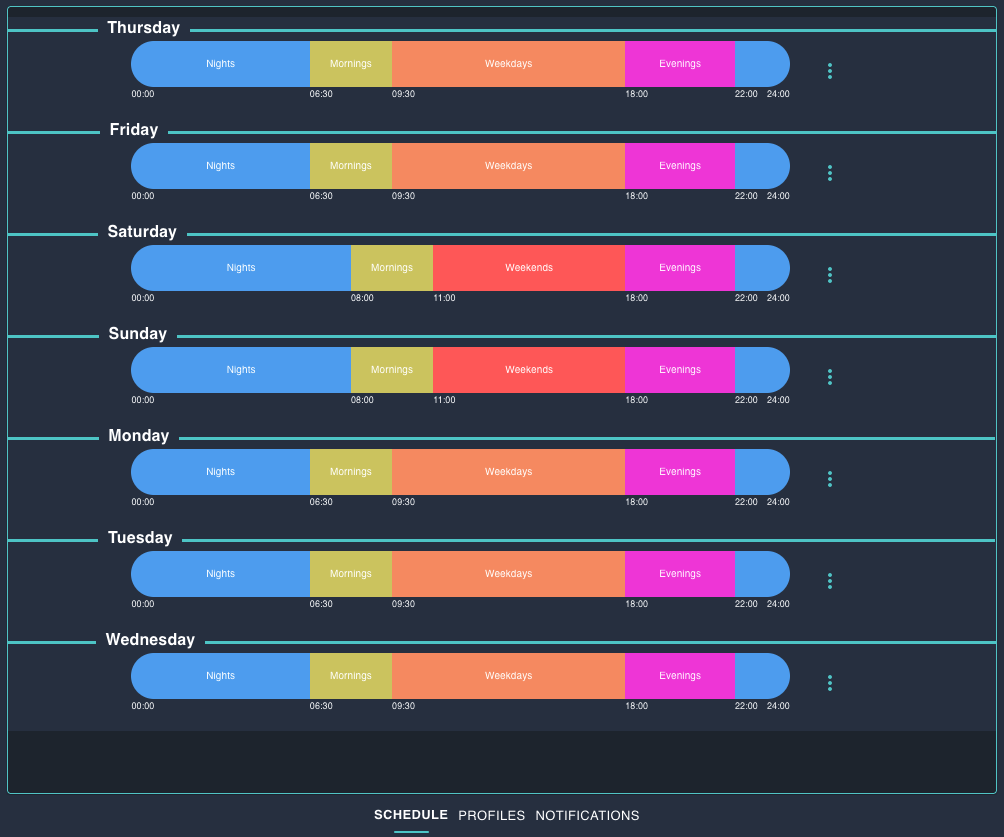}
		\caption{Schedule page.}
		\label{fig:schedule_page}
	\end{subfigure}
	\\\\
	\begin{subfigure}{0.5\textwidth}
		\centering
		\includegraphics[width=0.9\textwidth]{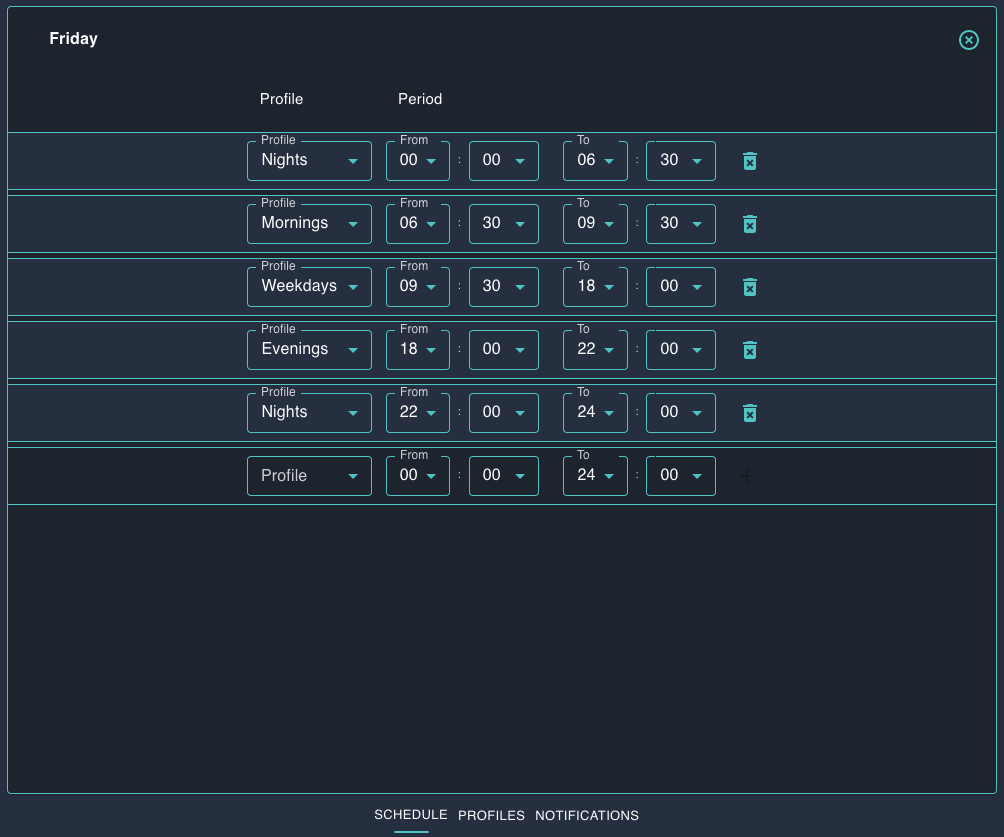}
		\caption{Edit overlay.}
		\label{fig:edit_overlay}
	\end{subfigure}%
	\begin{subfigure}{0.5\textwidth}
		\centering
		\includegraphics[width=0.9\textwidth]{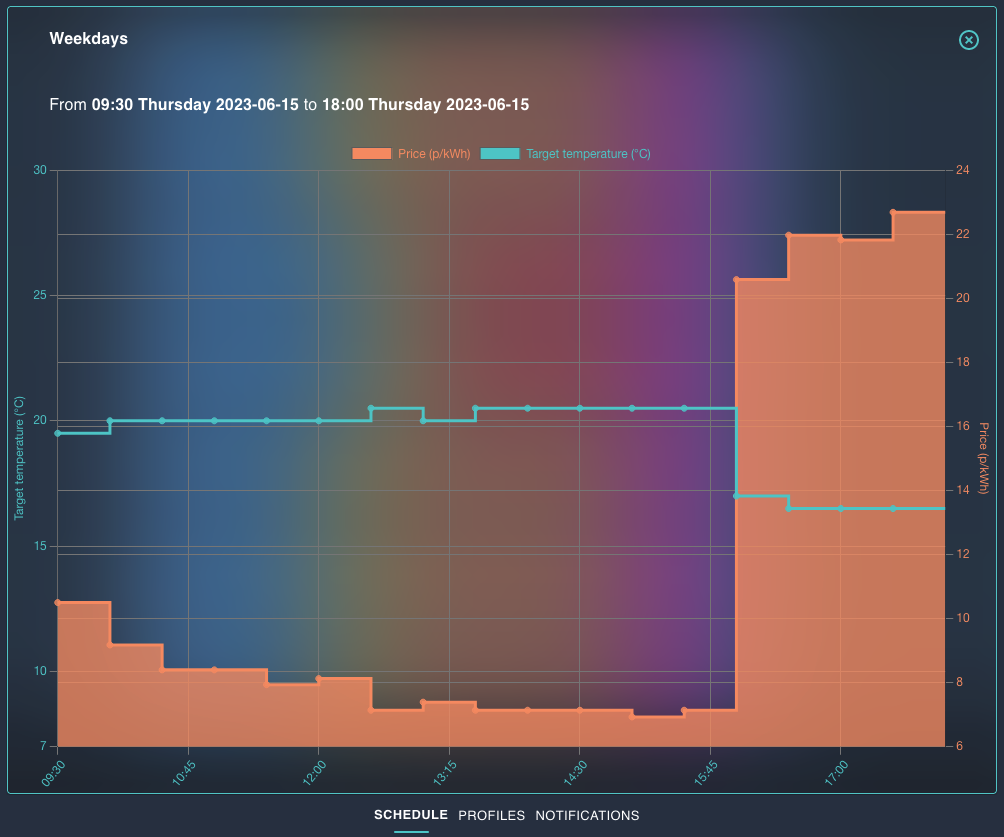}
		\caption{Timeslot overlay.}
		\label{fig:timeslot_overlay}
	\end{subfigure}
	\caption{Schedule page with overlays.}
\end{figure}

\begin{figure}[ptb]
	\begin{subfigure}{0.5\textwidth}
		\centering
		\includegraphics[width=0.9\textwidth]{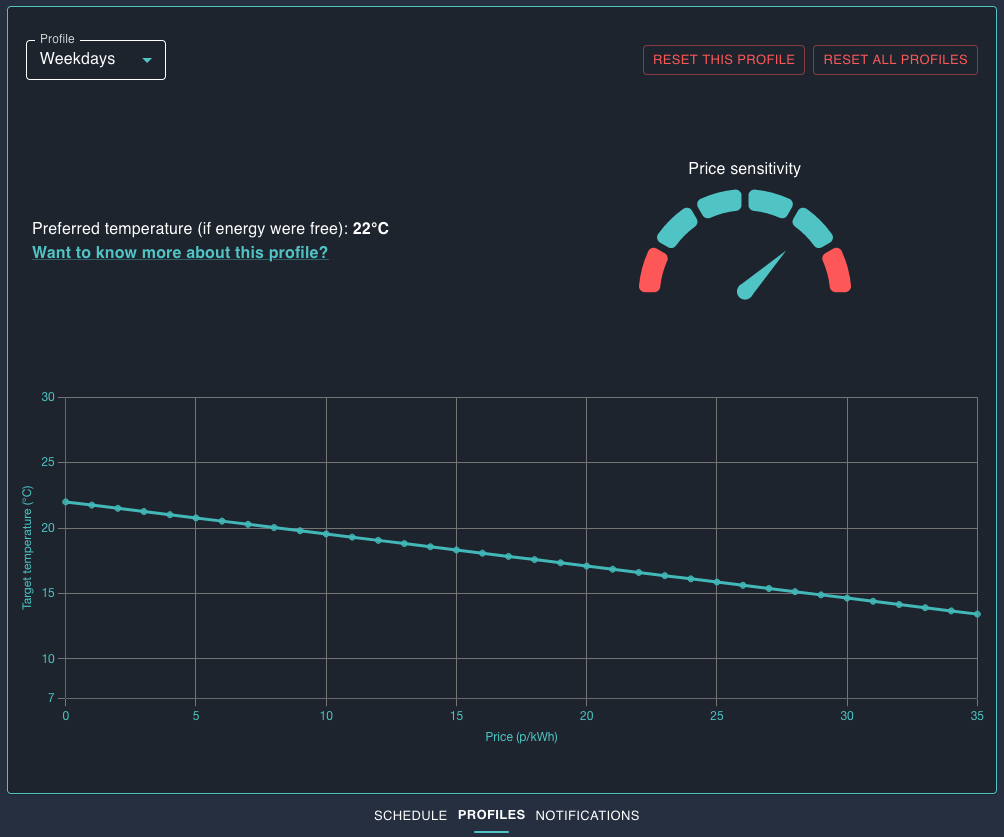}
		\caption{Profiles page.}
		\label{fig:profiles_page}
	\end{subfigure}%
	\begin{subfigure}{0.5\textwidth}
		\centering
		\includegraphics[width=0.9\textwidth]{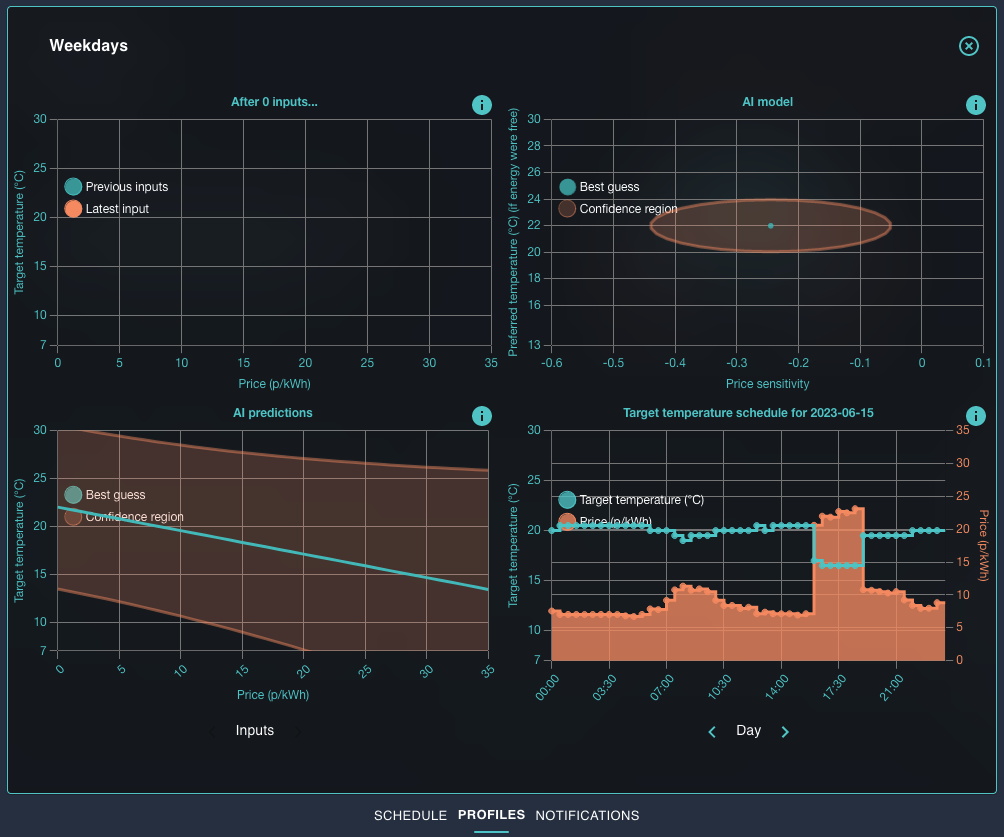}
		\caption{XAI overlay.}
		\label{fig:xai_overlay}
	\end{subfigure}
	\caption{Profiles page with overlay.}
\end{figure}

\subsubsection*{Schedule Page}
The \textbf{schedule page} (Figure~\ref{fig:schedule_page}) displays the user's current weekly schedule, comprised of an independent schedule for each day of the week.
The schedule follows the typical design used by programmable thermostats (e.g.\ EPH CP4) where each day is partitioned into a set of timeslots aligned to 15-minute intervals (e.g.\ 06:45--08:30).
In a programmable thermostat each timeslot would be allocated a setpoint.
In Squid each timeslot is allocated one of five \emph{profiles}, each having its own Bayesian model as defined in Section~\ref{sec:ai}.
The five profiles are colour-coded and given concrete meaning via predefined labels: \emph{Nights}, \emph{Mornings}, \emph{Weekdays}, \emph{Evenings}, and \emph{Weekends}.
When a profile is active according to the current schedule the system automatically adopts the setpoint specified by the profile's model for the current energy price.
User inputs only serve to update the active model. 
Note that the Netatmo hardware only supports setpoints in $\{ 7\degree\text{C}, 7.5\degree\text{C}, \dots, 30\degree\text{C} \}$, so all predicted setpoints are clipped to between 7$\degree$C to 30$\degree$C and rounded to the nearest 0.5$\degree$C.

All new user accounts are initialised with a default schedule, similar to Figure~\ref{fig:schedule_page}, which can then be customised by the user as follows.
Beside each day on the schedule page is a kebab menu with three options: \emph{Edit}, \emph{Copy}, and \emph{Clear}.
Selecting \emph{Edit} opens the \textbf{edit overlay} (Figure~\ref{fig:edit_overlay}), which consists of dropdown boxes and buttons allowing timeslots and allocated profiles to be customised for that day.
Selecting \emph{Copy} reveals a \emph{Paste} option beside other days, allowing timeslots and allocated profiles to be duplicated from one day to another.
Selecting \emph{Clear} resets the selected day to a single timeslot spanning 00:00--24:00 with the \emph{Nights} profile preallocated.
When any changes are made on the schedule page or edit overlay, \emph{Save} and \emph{Cancel} buttons appear, prompting the user to confirm or discard their changes.

Finally, selecting an individual timeslot on the schedule page opens the \textbf{timeslot overlay} (Figure~\ref{fig:timeslot_overlay}), which consists of a chart showing the upcoming energy prices for that timeslot along with the corresponding setpoints specified by the timeslot's model (i.e.\ these setpoints represent the upcoming setpoint schedule for that timeslot).
In this way the timeslot overlay allows users to inspect prices and setpoints for up to one week into the future.
If the selected timeslot is the current timeslot then today's prices are shown; the user can thus inspect historic prices and setpoints for earlier times, but only within the current timeslot.
The current day-of-week is always listed at the top of the schedule page to indicate that the underlying timeslot data is primarily future-directed.

\subsubsection*{Profiles Page}
The \textbf{profiles page} (Figure~\ref{fig:profiles_page}) allows users to inspect and/or reset any of the five profiles.
The page consists of three insights into the model of a selected profile:
a chart showing the mean of the model's predictive distribution (as in the fourth chart from Figure~\ref{fig:heatmap});
a text-based summary of the model's preferred temperature (if energy were free); and
the \emph{gauge}, which is a graphic visualising the model's price sensitivity.
When a user first navigates to the profiles page, the selected profile is always the active profile according to their current schedule.
Other profiles can then be selected using a dropdown menu located at the top-left of the page.
Profiles are reset by selecting one of the buttons located at the top-right of the page, including an option to reset only the selected profile and an option to reset all profiles simultaneously.
When either button is selected, \emph{Reset} and \emph{Cancel} buttons appear, prompting the user to confirm their choice.
When a profile is reset, its Bayesian model is replaced with the default model as defined in Section~\ref{sec:ai}, which is also the model used to initialise profiles for all new user accounts.

In the Bayesian model, price sensitivity is an (unbounded) dimensionless parameter, which may prove difficult for lay users to comprehend.
The gauge is intended to address this issue by interpreting the parameter on a six-point scale; it consists of four cyan-coloured segments, two red segments, and a dial that points to one of those segments based on the price sensitivity of the selected profile.
The two red segments appear on either side of the cyan segments.
The segments are determined by bucketing price sensitivity values into categories based on a lower bound $\mu_{s_{i}} = 0$ and an upper bound $\mu_{s_{i}} = \left( \mu_{b_{i}} - y_{\min} \right) / 35$ such that $y_{\min}$ is some low setpoint ($y_{\min} < \mu_{b_{i}}$).
This upper bound represents the point at which the model would select the low setpoint ($y_{\min}$) for a price equal to the maximum price permitted by the Agile tariff (35p/kWh).
Originally the low setpoint was defined as the minimum setpoint supported by the Netatmo hardware (7$\degree$C) but this was found to cause the gauge to be insufficiently discriminative under typical usage.
During the study the low setpoint was instead defined as 12.2$\degree$C based on observed gauge behaviour under typical usage.
The left red segment indicates that the price sensitivity is below the lower bound (i.e.\ negative price sensitivity), and the right red segment indicates that the price sensitivity is above the upper bound (i.e.\ contextually high price sensitivity).
The four cyan segments then indicate that the price sensitivity is within one of four equal-length intervals between the lower and upper bounds (i.e.\ 0\%--25\%, 25\%--50\%, and so on).
A colour gradient was not used for the gauge segments to avoid implying to users that a particular price sensitivity was objectively desirable (e.g.\ at the mid point).
For the special case where $y_{\min} \ge \mu_{b_{i}}$, the six-point scale was undefined and the gauge dial was hidden.

Finally, the profiles page includes a link labelled \emph{Want to know more about this profile?}
Selecting this link opens the \textbf{XAI overlay} (Figure~\ref{fig:xai_overlay}), which provides further insights into the model of the selected profile.
This overlay consists of four interactive charts equivalent to those proposed in Section~\ref{sec:xai}.
Buttons at the bottom-left of the overlay allow users to navigate through their history of inputs for this profile since its last reset; this causes all four charts to automatically update as illustrated in Figure~\ref{fig:interactive}.
In this way, navigating backwards fully (i.e.\ prior to any inputs) allows users to inspect the default model, while navigating forwards allows them to inspect how their model has evolved over time.
Buttons at the bottom-right also allow users to navigate through all available daily energy prices; since other charts are independent of this data, this only causes the fourth chart to update.
The unobtrusive positioning of these charts within an overlay rather than within a main page is intended to reduce cognitive load associated with day-to-day use of Squid while still making these detailed insights available to the interested user.
Tooltips beside each chart provide brief text-based descriptions (see Appendix) about how these charts should be interpreted by users.

\begin{figure}[tbp]
	\centering
	\begin{minipage}{0.5\textwidth}
		\centering
		\includegraphics[width=0.55\textwidth]{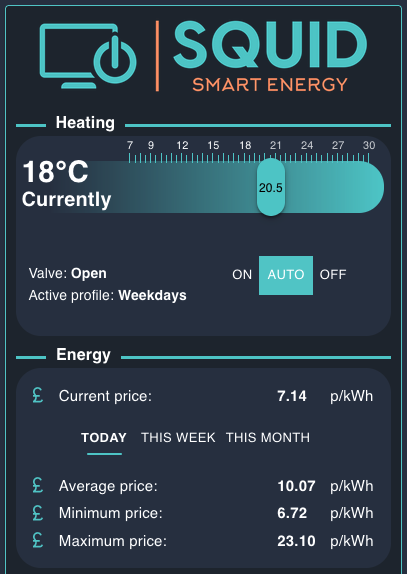}
		\caption{Quick Access panel.}
		\label{fig:quick_access_panel}
	\end{minipage}%
	\begin{minipage}{0.5\textwidth}
		\centering
		\includegraphics[width=0.9\textwidth]{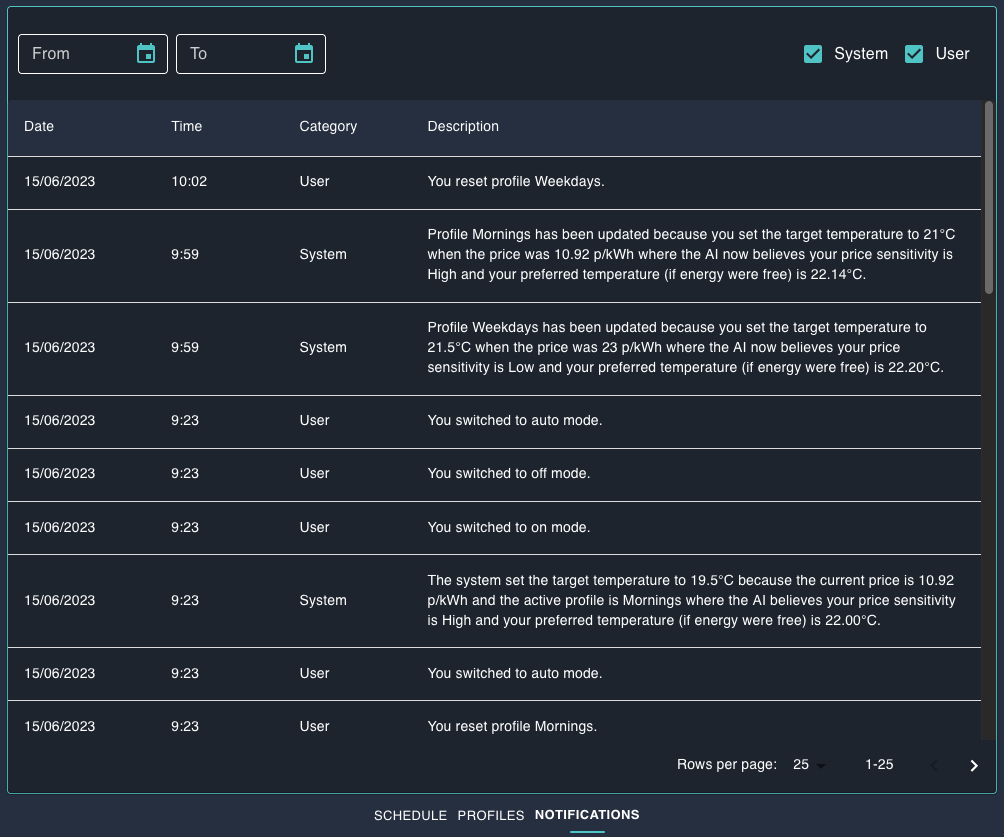}
		\caption{Notifications page.}
		\label{fig:notifications_page}
	\end{minipage}
\end{figure}

\subsubsection*{Quick Access Panel}
The \textbf{quick access panel} (Figure~\ref{fig:quick_access_panel}) represents the primary element of the user interface.
On larger displays (e.g.\ desktops) the panel is always visible and appears to the right of the user interface.
On smaller displays (e.g.\ mobile) the panel appears in an overlay accessed via a hamburger button at the top-left of the interface.
The panel is divided into two sections: \emph{Heating} (top) and \emph{Energy} (bottom).

The heating section consists of the current temperature indicator (top-left), setpoint controller (top-right), valve status and active profile indicators (bottom-left), and system mode controller (bottom-right).
The current temperature and valve status (open/closed) are as reported by the Netatmo hardware, while the active profile is the profile allocated to the current timeslot by the user's schedule.
The setpoint controller is an input slider that ranges from 7$\degree$C to 30$\degree$C in increments of 0.5$\degree$C (i.e.\ setpoints supported by the hardware).
The current setpoint is indicated by the slider handle.
Moving this handle allows the user to adjust the setpoint.
If the handle is moved, \emph{Yes} and \emph{Cancel} buttons appear, prompting the user to confirm their choice.
There are four system modes: \emph{on}, \emph{off}, \emph{auto}, and \emph{override}.
The system mode controller is a three-way toggle that allows the user to switch between these modes.
In \emph{on} mode, setpoints are temporarily disabled and the valve is set to remain open.
Likewise, in \emph{off} mode, setpoints are temporarily disabled and the valve is set to remain closed.
In \emph{auto} mode, the setpoint is automatically controlled by the system according to the active profile and current energy price.
Conversely, in \emph{override} mode, automatic setpoint control is temporarily disabled and the setpoint is instead based on the user's input.
Compared to SmartThermo~\cite{Alan:CHI:2016,Shann:AAMAS:2017}, on mode is equivalent to its boost option, while override mode is equivalent to its indirect learning design.
If either on or off mode is selected, the mode is enabled for a period of one hour, after which the system returns to auto mode.
If the setpoint is adjusted, the model is updated and override mode is enabled for a period of one hour, after which the system returns to auto mode.
If either on or off mode is enabled, the setpoint controller is disabled and its handle is hidden.
If override mode is enabled, both the setpoint controller and the auto option of the system mode controller are shown in orange, and the text \emph{Override active} appears below the current temperature.
If either on, off, or override mode is enabled, the mode expiry time is displayed below the system mode controller.
These modes may be terminated at any time by selecting the auto option on the system mode controller.
Override mode may also be terminated by selecting a \emph{Cancel} button that appears beside the system mode controller while in override mode.

Finally, the energy section consists of the current energy price indicator (top) and the price summary indicator (bottom).
The current energy price indicator simply displays the price for the current 30-minute slot.
The price summary indicator then displays summary statistics (min, max, average) for energy prices within one of three predefined ranges: the current day, current week, or current month.
The design of Squid assumes that a user's preferred setpoint will depend on the current energy price, so these summary statistics are intended to help users to interpret that price (e.g.\ whether it is high or low relative to other prices for the selected period).

\subsubsection*{Notifications}
The \textbf{notifications page} (Figure~\ref{fig:notifications_page}) provides a log of event notifications allowing users to inspect the complete history of events associated with their user account.
The notifications are displayed in a paginated table with four columns: date, time, category, and description.
The date and time columns indicate when the event occurred.
For the category column, events are partitioned into two types: user and system.
The user category includes four event types: user setpoint adjustments, system mode changes, profile resets, and schedule updates.
The system category includes two event types: auto setpoint adjustments (e.g.\ due to price changes) and profile updates.
Each event type has a notification template used to generate a text-based message shown in the description column.
These templates are available in the Appendix.
Notifications can be filtered by category using checkboxes located at the top-right of the notifications page.
Notifications can also be filtered by date using the date-range selector at the top-left of the page.
In addition to the notifications page, \textbf{flash messages} also appear briefly at the bottom of the user interface when certain events occur.
These flash messages occur for seven event types: system mode changes, price changes, active profile changes, setpoint changes, profile resets, and schedule updates.
Again, each event type has a template used to generate the message, which are available in the Appendix.

\subsection{AI Attacks}\label{sec:attacks}
Following the taxonomy of attacks on machine learning models proposed by Pitropakis et al.~\cite{Pitropakis:CSR:2019}, we focus on the three types of AI attacks for the field study:
\begin{description}
    \item[A-1 (Simple Poisoning)] Injection of malicious user inputs, mimicking an attacker that adjusts the setpoint so as to alter the model and affect future setpoint predictions.
    \item[A-2 (Complex Poisoning)] Same as A-1 attacks except that indicators of injected data are hidden, mimicking an attacker that has taken measures to conceal their inputs.
    \item[A-3 (Evasion)] Manipulation of energy prices, mimicking an attacker that affects future setpoint predictions without affecting the model itself.
\end{description}
All attacks were designed to cause the system to adopt a setpoint lower than it would otherwise have adopted.
A-1 attacks were simulated for each household by injecting 40 setpoint adjustments alternating between 7.5$\degree$C and 10$\degree$C over a period of 20 minutes.
A-2 attacks were simulated by injecting the same setpoint adjustments over a period of 60 minutes.
A-3 attacks were simulated for all households simultaneously by increasing prices with respect to a constant multiplier of $3$ for a period of five hours.
In the case of A-2 attacks, web interface indicators of attack inputs were hidden (e.g.\ notifications page entries, frames in the XAI overlay) but the \emph{effects} of those inputs were not hidden (e.g.\ on updated model parameters, on updated predictions).
No information was hidden in the case of A-1 and A-3 attacks.

All attacks had indicators that could be observed by the user, including web interface indicators and environmental indicators.
Web interface indicators included \emph{direct} indicators (i.e.\ evidence of attack inputs) and \emph{indirect} indicators (i.e.\ evidence of the effects of an attack).
Direct web interface indicators for A-1 attacks included anomalous entries on the notifications page, anomalous flash messages, and anomalous frames in the XAI overlay.
There were no direct web interface indicators for A-2 attacks.
The direct web interface indicator for A-3 attacks was anomalous price data (as shown on the quick access panel, timeslot overlay, notifications page, XAI overlay, and flash messages).
An indirect web interface indicator for all three attacks was anomalous predicted setpoints (as shown on the quick access panel, timeslot overlay, notifications page, and XAI overlay).
An additional indirect web interface indicator for A-1 and A-2 attacks was anomalous model parameters (as shown on the profiles page, notifications page, and XAI overlay).
Environmental indicators for all three attacks included a decrease in room temperature and the occurrence of motor sounds (buzzing) when the TRV actuator was triggered.
Alternating between setpoints of 7.5$\degree$C and 10$\degree$C during A-1 and A-2 attacks ensured that motor sounds would occur.
The effects of A-1 and A-2 attacks persisted until the affected model was reset, while the effects of A-3 attacks only persisted during the attack itself.
The primary mitigation action against the effects of previous A-1 and A-2 attacks was for the user to reset the affected model.
However, if the model was reset during an ongoing A-1 or A-2 attack then the newly reset model would be reinfected.
The primary mitigation action against A-3 attacks was to change the system to manual mode (on or off) and notify the software provider.
Several web interface indicators for previous attacks persisted regardless of mitigation action, including historic setpoints and model parameters (as shown on the notifications page) and historic price data (as shown on the quick access panel, timeslot overlay, notifications page, and XAI overlay).

\begin{table}[p]
	\centering
    \scriptsize
    \caption{Summary of households and participants.}
	\label{tbl:households}
	\begin{tabularx}{\textwidth}{C{0.01} C{1.32} C{1} C{0.9} C{0.6} C{0.71} C{1.2}}
		\toprule
		& \textbf{Household} & \textbf{Dwelling} & \textbf{Technology} & \textbf{Participant} & \textbf{Nationality} & \textbf{Occupation} \\
		\midrule
		\multirow[t]{2}{=}{1} & \multirow[t]{2}{=}{Parents\newline (with two children)} & \multirow[t]{2}{=}{House\newline (semi-detached)} & \multirow[t]{2}{=}{Smart meter} & Simone & France & Primary school teacher \\
		\cmidrule{5-7}
		&  &  &  & Theo & UK & Medical writer \\
		\midrule
		2 & Father\newline (with two children) & House\newline (detached) & Smart meter,\newline smart speaker,\newline smart doorbell & Isaad & UK/Morocco & Head of service delivery \\
		\midrule
		\multirow[t]{2}{=}{3} & \multirow[t]{2}{=}{Parents\newline (with two children)} & \multirow[t]{2}{=}{House\newline (semi-detached)} & \multirow[t]{2}{=}{Smart meter,\newline smart speaker} & Naadir & India & Town planning consultant \\
		\cmidrule{5-7}
		&  &  &  & Wiola & Germany & Academic in health \\
		\midrule
		\multirow[t]{2}{=}{4} & \multirow[t]{2}{=}{Parents\newline (with one child)} & \multirow[t]{2}{=}{House\newline (semi-detached)} & \multirow[t]{2}{=}{Smart meter} & Ernesto & Spain & Research manager \\
		\cmidrule{5-7}
		&  &  &  & Klara & Slovakia & Medical trainee \\
		\midrule
		5 &  Single occupant &  House\newline (terrace) &  Smart meter & Carrie & UK & Communications manager \\
		\midrule
		\multirow[t]{2}{=}{6} & \multirow[t]{2}{=}{Couple} & \multirow[t]{2}{=}{Flat\newline (one-bedroom)} & \multirow[t]{2}{=}{Smart meter,\newline smart doorbell} & Barış & Turkey & Civil engineer \\
		\cmidrule{5-7}
		&  &  &  & Maya & Turkey & Post-graduate student \\
		\midrule
		\multirow[t]{2}{=}{7} &  \multirow[t]{2}{=}{Mother and son} &  \multirow[t]{2}{=}{Flat\newline (two-bedroom)} &  \multirow[t]{2}{=}{--} & Sam & UK/Greece & Event worker \\
		\cmidrule{5-7}
		&  &  &  & Mara & Greece & Academic in education \\
		\midrule
		\multirow[t]{2}{=}{8} & \multirow[t]{2}{=}{Couple\newline (with three flatmates)} & \multirow[t]{2}{=}{Flat\newline (four-bedroom)} & \multirow[t]{2}{=}{Robot vacuum cleaner} & Kevin & Hong Kong & Marketing \\
		\cmidrule{5-7}
		&  &  &  & Ria & Taiwan & Sales \\
		\midrule
		\multirow[t]{2}{=}{9} & \multirow[t]{2}{=}{Couple} & \multirow[t]{2}{=}{Flat\newline (two-bedroom)} & \multirow[t]{2}{=}{--} & Ender & Turkey & CEO (accessibility start-up) \\
		\cmidrule{5-7}
		&  &  &  & Açelya & Turkey & CEO (education start-up) \\
		\midrule
		\multirow[t]{2}{=}{10} & \multirow[t]{2}{=}{Parents\newline (with two children)} & \multirow[t]{2}{=}{House\newline (semi-detached)} & \multirow[t]{2}{=}{Smart speaker} & Debora & Brazil & Head of research and development \\
		\cmidrule{5-7}
		&  &  &  & Aris & Greece & Academic in microbiology \\
		\bottomrule
	\end{tabularx}
\end{table}

\subsection{Recruitment}
Recruitment of users was conducted via community centres, social media, and snowball sampling of personal networks.
As an incentive, vouchers worth £120 were allocated for each household on successful completion of the study.
Vouchers worth £10 were also allocated for each AI attack successfully diagnosed (out of six in total).
Participation was sought to encompass a variety of household compositions and professions.
Households were screened to ensure they could commit to the proposed tasks and that no member of the household was vulnerable to shifts in room temperature.
In total, 36 households expressed interest, 11 households commenced the study, and 10 households (18 participants) successfully completed the study.
Details on the final households and participants are provided in Table~\ref{tbl:households}.
With one exception, all households were recruited via personal networks.
All households were located in England, with four in London and six in the south-west.
All households were middle class.
A majority of households were composed of either nuclear families or couples.
Dwellings were roughly balanced between houses and flats.
Only one household was a shared residence.
Participants included at most two adult members from each household.
Four participants were from the UK with the remainder from ten different countries in Europe, Asia, and South America.
With one exception, all participants had a minimum of third-level education.
All participants self-reported as confident technology users.
Two households self-reported as technology enthusiasts.
Barış \& Maya reported using Squid equally.
For other households, the first participant listed in Table~\ref{tbl:households} acted as lead user while the second acted as support user (e.g.\ only using Squid when the lead was unavailable).
Table~\ref{tbl:households} includes examples of home technology already in use by each household.

\subsection{Logistics \& Data Collection}
Ethics approval for the field study was obtained from each participating university.
All study participants were given an information sheet and signed a consent form.
All households commenced the study asynchronously between 15--29 January 2023.
The study then lasted seven weeks and was divided into two phases: \textbf{Phase~1} spanned Weeks 1--3 and focused on the use of Squid under normal conditions, while \textbf{Phase~2} spanned Weeks 4--7 and focused on the use of Squid under simulated AI attacks.
Application log data was captured for the duration of the field study in an anonymised form, including evidence of user activity (setpoint adjustments, mode changes, profile resets, schedule edits), user attention (page visits), and environmental conditions (temperature readings).
Any interviews were audio-recorded, transcribed, and anonymised.

\subsubsection*{Phase~1: Squid}
The study commenced for each household with \textbf{Home Visit~1}, which consisted of device installation, an interview, and a training session.
Following consultation with participants the valve was installed in a room reported to be in frequent use.
If there were two radiators in that room, participants were asked if they would agree to disable the second radiator for the duration of the study; two households did so.
Following installation was \textbf{Interview~1}, which was a semi-structured interview exploring household routines and prior experience with home technology.
This interview is not discussed in the remainder of this paper.
The visit concluded with \textbf{Training~1}, which was a 30-minute training session on the use of Squid, including how to adjust the setpoint, edit the profile schedule, reset profiles, and inspect the various charts and logs.
To reinforce the training session, four one-minute videos were also sent via email during Weeks 1--2 (two videos per week).
During Phase~1 participants were tasked with using Squid as normal.
In one instance (Simone \& Theo) this phase was reduced by one week due to household scheduling constraints.
To maintain engagement and offer technical support during this period, weekly 15-minute online check-ins were held with each household and documented by researchers as field notes.

\subsubsection*{Phase~2: AI Attacks}
The second phase commenced for each household with \textbf{Home Visit~2}, which consisted of another interview and training session.
\textbf{Interview~2} was a semi-structured interview that explored how participants were using Squid and their understanding of how it worked, including times when Squid was typically in use and whether participants were relying on any XAI features.
These interviews lasted 49 minutes on average (min: 35, max: 61, SD: 9).
The visit concluded with \textbf{Training~2}, which was a 30-minute training session on the AI attacks.
This included a live demonstration of an A-1 attack during which participants were shown how they might diagnose the attack (e.g.\ by inspecting notification logs) and mitigate its effect (e.g.\ by resetting the affected profile).
During Phase~2 participants were again tasked with using Squid as normal but were also tasked to monitor for AI attacks and respond to them as appropriate.
Households were then subjected to six simulated AI attacks (two of each type) at times unknown to them.
Attacks were scheduled to occur when households reported being most at home.
A-1 and A-2 attacks were scheduled such that at most one of either type occurred during any 24-hour period.
The order of attacks was randomised where possible, e.g.\ there was less flexibility in relation to A-3 attacks since they were designed to occur simultaneously for all households.
After each suspected attack, participants were asked to submit an online diary entry with contextual information, including who was present, what was perceived, and how they felt.
To maintain engagement and offer technical support, weekly 15-minute online check-ins were again held with each household and documented by researchers as field notes.

The study concluded with \textbf{Home Visit~3}, consisting of an exit interview and device removal.
\textbf{Interview~3} was a semi-structured interview that explored how participants responded to AI attacks, including indicators they relied on for diagnostics and their responsiveness in mitigating effects of the attacks.
These interviews lasted 63 minutes on average (min: 44, max: 99, SD: 16).
As a memory prompt, two online diary entries submitted by the household were discussed, giving participants an opportunity to reflect on their own experiences.
Finally, participants were reminded of all three attack types using visual aids, including their possible indicators and mitigations, giving participants an opportunity to reflect on specific challenges with each attack type.

\begin{table}[tbp]
	\centering
    \scriptsize
    \caption{Number of user activity events in Squid.}
	\label{tbl:user_activity}
	\begin{tabular}{l rrr | rrr | rrr}
		\toprule
		\multirow{2}{*}{\textbf{Type}} & \multicolumn{3}{l|}{\textbf{Phase~1}} & \multicolumn{3}{l|}{\textbf{Phase~2}} & \multicolumn{3}{l}{\textbf{Both}} \\
		& Mean & Median & SD & Mean & Median & SD & Mean & Median & SD \\
		\midrule
		Schedule edits & 5.4 & 4 & 5.9 & 0 & 0 & 0 & 5.4 & 4 & 5.9 \\
		Setpoint adjustments & 31.3 & 27.5 & 16.5 & 20.5 & 16 & 15.5 & 51.8 & 48 & 30.4 \\
		Mode changes (to on) & 1.2 & 0 & 1.7 & 0.5 & 0 & 1.6 & 1.7 & 0 & 2.5 \\
		Mode changes (to off) & 0.6 & 0 & 1 & 0.1 & 0 & 0.3 & 0.7 & 0.5 & 0.9 \\
		Mode changes (to auto) & 1.7 & 0.5 & 2.3 & 1.3 & 0.5 & 1.8 & 3 & 1.5 & 3.3 \\
		Profile resets (one profile) & 0.7 & 0.5 & 0.8 & 6.1 & 5 & 3.1 & 6.8 & 6 & 2.7 \\
		Profile resets (all profiles) & 0.4 & 0 & 1.3 & 0.2 & 0 & 0.4 & 0.6 & 0 & 1.6 \\
		\bottomrule
	\end{tabular}
\end{table}

\section{Results}\label{sec:results}
Table~\ref{tbl:user_activity} summarises the number of user activity events recorded by Squid during each phase of the study.
All schedule edits occurred in Phase~1, which is an observation that we would expect if household routines remained relatively fixed throughout.
Setpoint adjustments were by far the most common event, with users making at least one setpoint adjustment per day on average.
The number of mode changes was low overall and similar between Phases 1 and 2, which suggests that the mode controller was not much used, and either that mode changes were generally not used as a mitigation action for suspected A-3 attacks or simply that A-3 attacks were rarely suspected.
Comparing the number of profile resets between Phases 1 and 2 shows a sharp rise in Phase~2, which is unsurprising given that user training emphasised profile resets as a primary mitigation action for suspected A-1 and A-2 attacks.
During Interview~3, all participants reported that A-1 attacks were the easiest to detect, followed by A-2 attacks, with A-3 attacks being significantly more difficult.
Therefore, as mitigation actions, the number of mode changes and profile resets are consistent with how frequently participants were likely to have suspected the corresponding attack type(s).

In the remainder of this section we analyse Inteviews 2 and 3 in regard to the impact of XAI-related features on user comprehension and user satisfaction as well as their use as a tool in diagnosing AI attacks.
For this purpose, XAI-related features are any features that (i) explicitly reference either of the two model parameters or (ii) explicitly reference model inputs and/or model predictions.
The former applies to the profiles page, the second and third charts on the XAI overlay, and any system-category events on the notifications page (see Section~\ref{sec:software} for details).
The latter applies to the timeslot overlay, the quick access panel, the first and fourth charts on the XAI overlay, and certain user-category events on the notifications page (again see Section~\ref{sec:software}).
While the former may be more commonly accepted than the latter as being XAI-related, given that they expose the internals of the underlying glass-box models, the latter remains important in understanding how the AI feature makes use of those models.
Note that data on page visits was inadvertently polluted by researchers and thus is not reported in this section.

\subsection{User Comprehension}
The SmartThermo study was not focused on XAI, and largely predates the re-emergence of XAI seen in recent years.
The authors nonetheless analysed user comprehension of its AI feature, which they framed in terms of user mental models as demonstrated during exit interviews~\cite{Alan:CHI:2016}.
Results from that analysis offer an interesting companion to our own study.
Even by the end of that study, some SmartThermo users were demonstrating little curiosity in its AI feature: ``when we asked the users' opinion about what the thermostat was learning in the interviews, three users [out of 20] reported that they had not thought about it before.''
For the remaining users, the authors said that ``most participants appeared to have an understanding that [was] well-matched with the actual underpinnings of the
thermostat's learning feature'', e.g.\ one user described it as follows:
\begin{quote}
	As you input your set point changes according to the prices and then the system starts to understand what your views are of that cost I suppose.
	That is what you think is expensive and that is what you think is cheap, and then make changes.
\end{quote}
Crucially the authors also said that several users demonstated an incorrect mental model of the AI feature, with their understanding being that it learned user setpoint preferences as a function of time (without reference to price), e.g.\ according to one user:
\begin{quote}
	If I play a particular temperature as the setpoint and then click on save and learn, from what I understand is the system will take this reading to consideration for whether to turn the boiler on or off but at the same time try to see that at this particular time of the day, whether it's weekday or weekend and then try to replicate that during other days.
\end{quote}
The authors did not report exactly how many users demonstrated this incorrect mental model but said that it ``was more prevalent among the indirect learning thermostat users than among the users of the direct learning thermostat''.
This suggests that the number of users with this incorrect mental model was non-neglible.
Moreover, the fact that it was more prevalent among indirect learning users makes it of particular relevance to our study, given that the indirect learning design was the design replicated in Squid.
According to the authors, no user reported familiarity with existing consumer smart thermstats, including those that do seek to learn user setpoint preferences as a function of time (e.g.\ Nest Learning Thermostat), so they discounted the possibility that users may have been biased by prior experience.

Reflecting on our interviews, we found no evidence that any participant developed an incorrect mental model of the AI feature in Squid, including one similar to that seen in the SmartThermo study.
Of course, Squid and SmartThermo were not part of the same study so we cannot draw concrete conclusions about why our results were different.
Nonetheless, our results are consistent with a hypothesis that design changes in Squid would have a positive effect on user comprehension compared to SmartThermo.
These design changes include (but are not limited to) the addition of multiple Bayesian models as outlined in Section~\ref{sec:software} and the addition of some XAI-related features as outlined earlier in Section~\ref{sec:results}.
While we introduced temporality into the AI feature via multiple models, there is no evidence that any participant believed that the models themselves learned user setpoint preferences as a function of time.
For example, Ernesto and Theo both gave accurate explanations regarding the model itself, and although Theo expressed some confusion regarding the impact of having multiple models, Simone gave accurate explanations of both:
\begin{quote}
	\textbf{Ernesto:} The AI will try to adjust, to reduce the temperature, but not a lot, in order to save some money, but no--- medium temperature, something like that.
	And it's for us to really tell the AI, ``This is too cold,'' or, ``This is too hot,'' at these times. You look at the price, say, ``I can afford to pay a bit more if I increase the temperature.''
	The AI says, ``This is too expensive,'' let's put it down.
	So it's kind of like using the data and then us, what we want to teach it, to ensure that we get an agreement we can both---
\end{quote}
\quotesep
\begin{quote}
	\textbf{Theo:} Yeah, so my thought was that obviously when the price was high the algorithm was trying to lower the temperature a little bit in an effort to save money. But if then we went in and responded and said, ``No, it's too cold'' and we forced it up, presumably it then learned and then didn't do that again, or did it less.
	And then I'm not sure whether that's then applied to all the profiles or which profile it then applies it to or whether you had to do it every single time for each profile.

	\textbf{Simone:} I think you do.
	I agree with what Theo just said.
	My understanding was that you need to go often enough at least once inside each of the profiles so that the AI knows that on a Monday morning or weekday or weekend this is your target temperature.
	And if you do it on a weekday on a Monday it's going to build it for every weekday and you also need to do it on the weekend so every profile for every day the AI memorises what your preferences are.
	And exactly like Theo said, it's depending on--- it's offsetting the price increases so building it for you so you can save money without having to play around with your own thermostat.
\end{quote}
While participants mostly understood the inputs, outputs, and predictions of the AI feature, there is no particular evidence that they understood how the model was updated.
We will discuss this further in Section~\ref{sec:satisfaction} in the context of specific XAI-related features.

\subsection{User Satisfaction (Phase 1)}\label{sec:satisfaction}

\begin{figure}[tbp]
	\centering
	\includegraphics[width=0.65\textwidth]{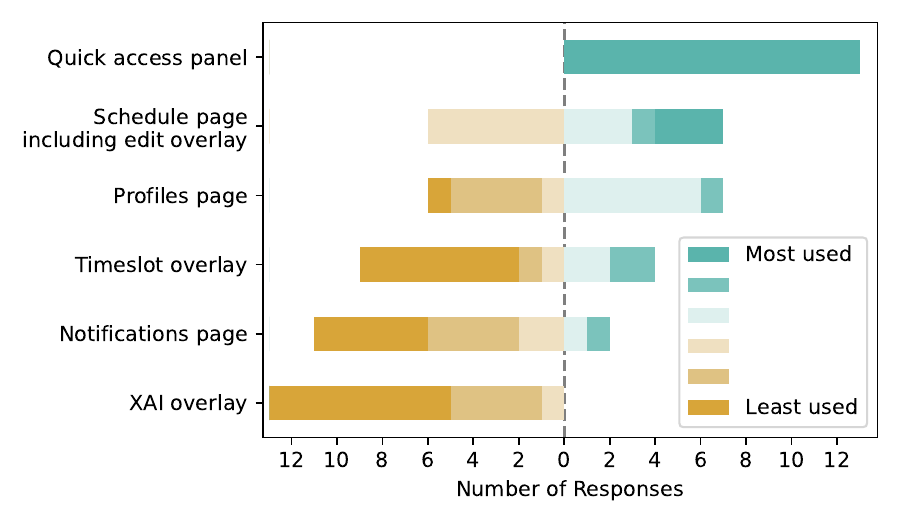}
	\caption{Feature rankings reported at the end of Phase~1 with joint rankings from five households and individual rankings from the remaining eight participants.}
	\label{fig:rankings}
\end{figure}

At the conclusion of Phase~1, during Interview~2, participants were asked to rank six features from the web interface on a six-point scale from most-used~(6) to least-used~(1).
Five couples provided joint rankings, while the remaining eight participants provided rankings individually, giving 13 rankings in total.
The frequency of responses for each feature is shown in Figure~\ref{fig:rankings}.
Unsurprisingly the quick access panel was ranked by all participants as their most-used feature (all: 6).
This panel was displayed prominently in the interface and provided key system controls, including setpoint adjustments and mode changes.
The schedule page including edit overlay tended to be ranked as the next most-used feature (mean: 4.08, median: 4, mode: 3).
This page provided schedule-edit controls, which was shown in Table~\ref{tbl:user_activity} to have been moderately used during Phase~1; it also served as homepage for the interface.
The profiles page tended to be ranked slightly lower than the schedule page (mean: 3.15, median \& mode: 4).
This page provided profile-reset controls, which was shown in Table~\ref{tbl:user_activity} to have been rarely used during Phase~1.
The timeslot overlay tended to be ranked next (mean: 2.31, median \& mode: 1), placing it among the less-used features.
This overlay provided information on energy prices and the corresponding setpoint schedule but had no system controls and was accessed indirectly via links on the schedule page.
The notifications page tended to be ranked slightly lower than the timeslot overlay (mean: 2.15, median: 2, mode: 1).
This page provided information about system events but again had no system controls.
Finally, the XAI overlay tended to be ranked as the least-used feature overall and was ranked among the less-used features by all users (mean: 1.46, median \& mode: 1).
This overlay provided information on the AI models but had no system controls and was accessed indirectly via a link on the profiles page.
A clear pattern emerges: features that included system controls were ranked higher than those that were purely informative; features that were displayed prominently were ranked higher than those that were displayed unobtrusively; and features that included frequently-used controls were ranked higher than those that only included infrequently-used controls.

\subsubsection*{Quick access panel}
As mentioned, the quick access panel was reported as the most-used feature during Phase~1, while setpoint adjustments (made via the quick access panel) were found to be the most common user activity event.
Most participants made comments that suggested they understood the AI feature was learning from these setpoint adjustments, e.g.:
\begin{quote}
	\textbf{Debora:} It learns that that's the temperature that I want.
\end{quote}
However, several users made comments implying that while they understood the functionality on a theoretical level, they were not necessarily conscious of its consequenes when they were making setpoint adjustments during day-to-day use of Squid:
\begin{quote}
	\textbf{Carrie:} Well, it is registering what I am doing. I haven't really thought about that.
	It is registering what I am doing. And I guess it recalibrates a bit.
\end{quote}
\quotesep
\begin{quote}
	\textbf{Theo:} I think sometimes you can think you're in the moment now and
	you don't register that that's then going to apply throughout maybe.
\end{quote}
SmartThermo made the consequence of setpoint adjustments more explicit by labelling its confirmation dialog button as \emph{Save \& Learn}, whereas Squid labelled this button simply as \emph{Save}.
In hindsight, the more explicit form may have helped to remind users of the consequence of their action and thus may have mitigated the issue raised by Theo.

\subsubsection*{Timeslot overlay}
Ender and Açelya both described the timeslot overlay as ``understandable''.
Sam was able to explain the timeslot overlay succinctly but did so without linking it to the broader AI feature, whereas Maya explained it by linking explicitly to user preferences:
\begin{quote}
	\textbf{Sam:} It's just adjusting [the setpoint] to the price. That's all it is doing.
\end{quote}
\quotesep
\begin{quote}
	\textbf{Maya:} I think it is very clear what it says, this the price, this is the temperature and when the price is really high the AI tends to lower the temperature a little bit but in accordance with your desired temperature, so it goes back up.
\end{quote}
In terms of usage, Debora said that she relied on the timeslot overlay to inspect her upcoming setpoint schedule, which was indeed its intended function:
\begin{quote}
	\textbf{Debora:} It's more for me to see how I'm doing.
	Because I increase it, so it changes it automatically.
	So I'm actually trying to see again my own trends on it.
\end{quote}

\subsubsection*{Profiles page}
Users were generally positive about the chart on the profiles page and expressed clear understanding of what it was intended to convey:
\begin{quote}
	\textbf{Simone:} So here the first thing I look out is that the graph, the line is going down.
	If it's going down that means that the target temperature goes down when the price goes up, so it's what you would expect from using the Squid is the sensible way of looking at it.
\end{quote}
\quotesep
\begin{quote}
	\textbf{Debora:} Well, because there is temperature versus price, so it should have an inversely-proportionate relationship, yes?
	If what you are targeting is to save money, that's how it should be.
\end{quote}
A notable exception was Isaad, who was altogether more negative about its usefulness:
\begin{quote}
	\textbf{Isaad:} What is it telling me? So for me, everything is--- If I want a piece of information, it needs to be something actionable. A piece of information that doesn't drive a piece of action is pointless information. So I'm like, what does that tell me? It doesn't tell me to do anything or change my behaviour, or even tell me anything about my existing behaviour. I'm just like, okay, it's going down, that's it.
\end{quote}
No user made an explicit connection between the linear format of the chart and the two model parameters displayed at the top of the page, i.e.\ where the \emph{preferred temperature (if energy were free)} value is the $y$-intercept of the chart and the \emph{price sensitivity} gauge is a visualisation of the slope.
While some users reported that the chosen terminology was intuitive, others implied that this was not necessarily the case:
\begin{quote}
	\textbf{Ernesto:} But it's a nice [logical] thing. Free. The real ideal temperature, there was a
	cost, the lower it will be, obviously.
\end{quote}
\quotesep
\begin{quote}
	\textbf{Theo:} Maybe it's just like a--- the way it's displayed or the langauge like ``Squid's calculated this would be your preferred temperature'', if there was something like that. Or ``Based on your behaviour this is your price sensitivity'', just stuff like that to guide you a little bit more.
\end{quote}
Despite Squid being presented as an application that uses AI to \emph{learn} user preferences, several users appeared to lose sight of this when they questioned how the system was able to predict their preferred temperature (if energy were free) even though they had never faced such a situation before:
\begin{quote}
	\textbf{Simone:} So aside from what I've just seen there, when I've been before I have seen this, the preferred temperature if energy was--- and
	it was something like 20 something degrees \dots but I thought, ``How is that calculated?'' because I always pick 19. Of course, there's always price in there, so how do you calculate my preference if energy was free?
\end{quote}
\quotesep
\begin{quote}
	\textbf{Sam:} How can they make that assumption? Because they haven't seen, I mean I've never put it at 22, 23 when it was [free]
\end{quote}
Again it seems that the chosen terminology may have been a factor in this confusion.

The gauge divided opinion and was generally not well-understood.
For example, while Debora said the gauge was ``very interesting'', Sam and Mara both said it was not understandable.
The concept of price sensitivity as a kind of measurable character trait seemed to engage users, but in some cases the learned value conflicted with the user's broader sense of self, e.g.\ commenting on a profile with a low price sensitivity value:
\begin{quote}
	\textbf{Ender:} In fact I am not that less price sensitive, I am just more price sensitive and I am just checking for offers, promotions and everything.
\end{quote}
Ernesto understood that the gauge and chart were based on the same information but he preferred the latter and did not understand exactly how the two were related:
\begin{quote}
	\textbf{Ernesto:} The graph is more informative and takes the information out of the gauge.
\end{quote}
\quotesep
\begin{quote}
	\textbf{Ernesto:} But I don't really know get a lot what happens when it's in here, how that affects the graph and everything else.
\end{quote}
The design of the gauge was intended to describe the current price sensitivity value in terms of a \emph{typical} range, with the teal segments indicating typical values and the red segments indicating atypical values.
This representation seems to have been successful:
\begin{quote}
	\textbf{Simone:} And then looking at the gauge as well, that you should be in the green.
\end{quote}
\quotesep
\begin{quote}
	\textbf{Barış:} It is good to be on the blue region.
\end{quote}
In one instance Carrie noticed that the gauge was pointing to a red segment and provided a reasonable explanation about why this may have been the case:
\begin{quote}
	\textbf{Carrie:} \dots generally you would expect it to be within the green ones.
	\dots
	But also once it did go right up into the red I thought, ``Ooh, why has it done that?''
\end{quote}
\quotesep
\begin{quote}
	\textbf{Carrie:} Maybe it was actually the evening, so where the price was low, and I set the temperature lower.
	I think it was that, yeah.
\end{quote}
However, the semantics of the red segments were generally unclear:
\begin{quote}
	\textbf{Maya:} I think if it goes that side it means that we are highly sensitive and the other side we don't really care about, or is it the
	opposite?
\end{quote}
\quotesep
\begin{quote}
	\textbf{Sam:} I'm in the red now but it doesn't tell me why, what does that mean.
\end{quote}
In some cases, incorrect semantics were assigned to these red segments by users:
\begin{quote}
	\textbf{Barış:} If we go to left, red, that means we're keeping cold.
\end{quote}
\quotesep
\begin{quote}
	\textbf{Aris:} If it goes to the red, you are spending everything.
\end{quote}
As mentioned in Section~\ref{sec:software}, the gauge was also designed to avoid implying to users that a particular price sentivity value in the typical range was objectively desirable.
While this design choice was largely successful in its aim, the lack of an objectively desirable price sensitivity value was a source of frustration for Isaad:
\begin{quote}
	\textbf{Isaad:} It's just trying to stay in the middle.
	It doesn't tell me anything. That only has a meaning if there's a target, if there's a limit, if
	there's something you're aiming for. This is just, yes, indicative, but indicative of what I don't know, and, again, it doesn't drive me to do anything.
\end{quote}
This comment echoes the earlier comment from Isaad that the chart on the profiles page did not provide actionable information.
However, these two comments were outliers; most users seemed satisfied that the profiles page described what the system had learned about their preferences, and that this information was implicitly actionable, i.e.\ if the user was unsatisified with the current learned preferences, they could intervene by providing more inputs and/or by resetting the profile.
Making this actionability more explicit on the profiles page may have mitigated the issues raised by Isaad.

The profiles page allowed users to inspect (and thus compare) any profile by selecting the corresponding option on a dropdown menu located at the top of page.
Naadir felt this option was not sufficiently prominent, causing him to largely overlook the feature:
\begin{quote}
	\textbf{Naadir:} I admit that this was not natural to come to me to change which profile it was because I was so focused on that that to go and look at the other ones I wouldn't necessarily do that.
	And if it was maybe more out in front of you then you're more likely to go click it. Because it had hidden at a dropdown bar, I was [not] focused on that.
\end{quote}
Isaad on the other hand suggested that the profiles page could be improved by allowing different profiles to be explicitly compared on the same page:
\begin{quote}
	\textbf{Isaad:} So, for me, it'd be more interesting to compare the different profiles to each other than it would to be looking at the actual profile.
\end{quote}
This suggestion by Isaad may have served to better contextualise the two model parameters for users, and would likely have avoided the issue raised by Naadir.

\subsubsection*{Notifications page}
Most users framed the notifications page as a traditional log file.
Sam described the entries as a representation of his ``history''.
Debora, Aris, and Carrie all described the notifications page as ``useful''.
Kevin and Ernesto both noted that the description column included detailed information about the AI control mechanism, although Ernesto complained of information-overload:
\begin{quote}
	\textbf{Kevin:} And then the algorithm will just make the changes based on our preferences, so the algorithm just will make any changes, and then it would just save it here as, kind of, like, a record.
\end{quote}
\quotesep
\begin{quote}
	\textbf{Ernesto:} But it records everything. Every change that the AI does, it's
	recorded there.
	\dots
	It's too much information, and some of it says the same, about the valuation, which is what [manages] the figures, the temperature.
	And you have so many of them that, to know which one can be right or wrong, you may be lost in the amount of information.
\end{quote}
Echoing what Ernesto said about correctness, Sam and Isaad both noted that the notification page would be used primarily to diagnose system issues:
\begin{quote}
	\textbf{Sam:} Yes, I would find it useful if something was not functioning and then I'd go look at it and it's saying that I actually did the thing but it's not working.
\end{quote}
\quotesep
\begin{quote}
	\textbf{Isaad:} This is everything that's happened, from every change that you request, to any tweak.
	So if anything's happened that you haven't done then you can ask yourself a question.
	But again, you go to this by exception.
	If something goes wrong, then you go to this to find out what went wrong.	
\end{quote}
As an example, Carrie and Ender both used the notification page to check if the system was functioning according to their expectations, focusing on the description column:
\begin{quote}
	\textbf{Carrie:} I generally go in just to look at basically the top one.
	I might skim down two or three or four below, just to see what the AI has been doing when I have not--- Just to see if it says something different from what I was expecting maybe.
	\dots
	I think the thing that is interesting me there is the temperature if energy were free.
	I am keeping an eye on that really.
	But also then it says the price was only 8p when I set it down.
\end{quote}
\quotesep
\begin{quote}
	\textbf{Ender:} Especially I checked it in the morning, so when I just first used the app, I checked what happened during the night.
	\dots
	for example in the morning it changed to 21 and we didn't adjust it but maybe it could have saved something for us.
	I'm just checking is it working kind of information, so is it really doing something with the valve \dots
\end{quote}
Conversely, Naadir and Wiola used the notifications page to check whether their inputs were being recorded by the system but did not use it to understand system behaviour:
\begin{quote}
	\textbf{Naadir:} Well, mainly when I flicked it up just to go and see if it was logged, that was it.
	I didn't go and check and see if the system was recording things, or at least not that I recall, I just--- maybe but to check.

	\textbf{Wiola:} Yeah, when we first started to use it you went back and checked that, whatever changes we had made were on there---
\end{quote}

\subsubsection*{XAI overlay}
Among the four charts on the XAI overlay, the fourth chart was frequently cited as the most useful and intuitive.
Carrie and Ria both described the fourth chart as ``easy to understand'', while Ender said it was ``understandable''.
Theo said the fourth chart ``makes the most sense'', which was a sentiment echoed by Simone:
\begin{quote}
	\textbf{Simone:} Yeah, I agree actually, I was just going to say that. That is interesting, I think this tells me what it does actually, it tells me how the AI is reducing the temperature when it hits a certain price.
\end{quote}
Isaad preferred the fourth chart because it was most clearly tied to real-world effects:
\begin{quote}
	\textbf{Isaad:} It's the only one that actually means something to me. If that's the target temperature, that's how I'm going to feel it. So that's the only one that actually really affects me.
\end{quote}
Wiola felt that the fourth chart was intuitive and she demonstrated good understanding of what the chart was intended to convey:
\begin{quote}
	\textbf{Wiola:} The fourth one, so you have the time on the $x$ axis see how the time occurs all day and then you have the two $y$ axis basically for temperature on the left, see how the temperature was changed over the days across the time. And then on the right is the price I guess and other changed over time and then you can see how they relate to each other as the right times. But I think that is very intuitive, I like that one.
\end{quote}
The fourth chart and the chart on the timeslot overlay are effectively the same chart, with the main difference being the data on the $x$-axis: the former uses price data for the selected timeslot while the latter uses price data for the selected day.
Despite this equivalence, no participant made a connection between the two charts.

At the opposite end, the second chart was frequently cited as the most confusing and frustrating.
Simone reported that the second chart confused her while Naadir said that he had ``no clue what it's trying to show''.
Likewise, Ernesto reported that he did not understand the second chart, and Ria reported that it was the least informative.
While Carrie demonstrated reasonable intuition about the size of the confidence region, she felt she did not understand the meaning of its shape and/or orientation:  
\begin{quote}
	\textbf{Carrie:} With the bubble one, I have seen that has sometimes gone shorter and fatter. Shorter rather than [long], which I assume is getting more accurate. I assume. But I don't understand the tilt or anything.
\end{quote}
Ender felt that the price sensitivity terminology was unclear and struggled to connect it to the specific numbers on the corresponding axis:
\begin{quote}
	\textbf{Ender:} I don't understand, for example, the top-right one, I don't understand that, it is okay but I don't understand what $-3.5$ means depending on the thing for me, I don't get that one.
\end{quote}
Only Barıs and Maya made a connection between the second chart and the gauge, and while Maya demonstrated reasonable understanding of this connection at first, Barış misunderstood the connection, and this caused Maya to doubt herself:
\begin{quote}
	\textbf{Maya:} The gauge, I think it is the graph version of the gauge, right? It is price sensitivity going a little bit left-hand side on the gauge, so from middle to the minus side.

	\textbf{Barış:} I agree, if you go out it means you are out of the circle. If you go to the red zone---

	\textbf{Maya:} Yes, if you are out of the ellipses side you are in the red zone maybe, I'm not sure.
\end{quote}
The true connection between the second chart and the gauge can be seen in the definition of the gauge segments from Section~\ref{sec:software}: a value of zero on the $x$-axis ($\mu_{s_{i}} = 0$) represents the dividing line between the left red segment and the five remaining segments.
In hindsight it may have been preferable to indicate the gauge segments on the chart itself, perhaps by labelling the $x$-axis with the segment labels shown on the notifications page (negative, very low, etc.---see Appendix).
Of course, this would have come at the cost of increased visual complexity on the XAI overlay.

Most users reported that the first and third charts were useful and intuitive, but this opinion was not uniform and they were generally not cited as frequently as the fourth chart.
Carrie described the first chart as ``interesting'' because it shows ``interaction with the AI''.
Aris cited the first chart as the most useful.
Theo and Wiola both demonstated clear understanding of what the first chart was intended to convey, e.g.:
\begin{quote}
	\textbf{Theo:} I think isn't that that's the input just like we've told it that we chose that temperature and when we said that, the cost was what is on the $x$ axis. So we said 17 degrees, which is the $y$ there and at that point when we tapped it it costs whatever it was, 12 pence per kil--- and so that's a factor that's gone into the algorithm that we like. We're willing to have such and such temperature for such and such price.
\end{quote}
Conversely, Simone said that the first chart confused her (a sentiment she also expressed regarding the second chart), seemingly in relation to its interactiveness:
\begin{quote}
	\textbf{Simone:} Oh, so does that mean you can have after three inputs, after four? Why is it after two?
\end{quote}
Unusually, Kevin cited the first chart as least informative, although he did not elaborate.
He also cited the third chart as most informative:
\begin{quote}
	\textbf{Kevin:} \dots for me it's quite useful to know what the AI will predict on the temperature, based on my previous preferences. So that's why I think this one for me. Because if I know how the AI will work, I am quite interested in it, so that's why I think it's quite informative for me.
\end{quote}
Similar to the first and fourth charts, Wiola demonstrated reasonable understanding of what the third chart was intended to convey:
\begin{quote}
	\textbf{Wiola:} This is how it adjusted based on so if it's really--- if the price is really low then the temperature--- it would increase the temperature? That's I guess for people who want to make sure that when the price goes higher the temperature is automatically going down. And then, yeah, the confidence entered all around it so it's not always exactly--- right.
\end{quote}
Only Wiola and Naadir explicitly commented on the linearity of the AI model shown in the third chart, and in fact their comments imply that they disagree with the assumption that underpines both Squid and SmartThermo (see Section~\ref{sec:discussion} for a discussion):
\begin{quote}
	\textbf{Wiola:} You don't want it linearly, you want it as a---

	\textbf{Naadir:} Yeah, it'd be a curve, it would be--- or just hit a floor and that would be straight across.
\end{quote}
The third chart on the XAI overlay and the chart on the profiles page are effectively the same chart, with the main difference being that the former includes a confidence region.
Despite this equivalence, no participant made a connection between the two charts.

Tooltips beside the second and third chart describe the confidence regions as representations of ``uncertainty over the best guess'' such that ``a larger confidence region means more uncertainty'' (see Appendix for full text).
Nonetheless, the confidence regions were poorly understood in general and were frequently cited as the most confusing aspects of the XAI overlay.
For example, Ernesto felt they were unintuitive:
\begin{quote}
	\textbf{Ernesto:} It's difficult. I look at it and I didn't understand exactly what you were saying about the confidence region. What do you mean, the confidence region? Where the AI thinks that my temperature is going to be? Or--- No, I didn't really understand that one.
\end{quote}
Barış and Maya both misinterpreted the confidence region on the third chart:
\begin{quote}
	\textbf{Barış:} According to inputs gets a line and puts some safety zones, so up and down, and if we are between these lines that is okay.
\end{quote}
\quotesep
\begin{quote}
	\textbf{Maya:} I think the orange is our comfort zone, it is our desired minimum and maximum thing and the middle line is the optimum line.
\end{quote}
Comparing the confidence regions on both charts, Carrie felt more comfortable with the confidence region on the third chart:
\begin{quote}
	\textbf{Carrie:} Actually, I have to admit, the confidence one I don't understand terribly well. I am not a data person. Actually, on one of the tip videos the bubble was upright, was vertical, and I wondered whether that is actually what it should be like if it is perfect, that is where it should be. But I don't quite understand how that works.
	The AI prediction, so the confidence range, that is easier, because I know the shape shifts. That is easier.
\end{quote}
Simone and Naadir both expressed surprise at the (relatively large) size of the confidence regions, implying that they understood what the regions were intended to convey, e.g.:
\begin{quote}
	\textbf{Naadir:} I'm surprised, I guess what I was surprised about is how wider the confidence in that particular graph, but I don't know, that would be the only thing. Because on the left hand side it's got almost a 10 degree variant which is quite wide and 15 degrees is quite cold and 25 degrees is quite warm.
\end{quote}
Unusually, Mara felt that the confidence regions were useful and likewise demonstrated reasonable understanding:
\begin{quote}
	\textbf{Mara:} That's really useful I think because I would want--- It calculates some price sensitivity to whatever it is, $-0.2$ and $-20$\%, and then that's at 22, which was your wishful thinking.
	\dots
	That's useful, yes. It kind of, it tells you how confident the system is, right, the confidence region.
\end{quote}
In general, the confidence regions were the least successful aspect of the XAI overlay.

As noted already, the XAI overlay was generally not used much during Phase~1.
Some users complained that the XAI overlay required too much effort to understand, while others complained of information-overload.
Ernesto and Açelya both felt that the effort required to understand the XAI overlay was not worth any potential benefits:
\begin{quote}
	\textbf{Ernesto:} Yes, that's why the other one, those ones there, are difficult to actually understand. What do you mean with this? If you had the time, I would go and try to understand it and say, ``What did you mean by this?'' Because having a graph there that means nothing to you is not comfortable.
\end{quote}
\quotesep
\begin{quote}
	\textbf{Açelya:} I don't want to spend too much time to understand, for example. I can if I need to of course, I will look at and I will try to understand but for making my life easier that will be much more easy and catchy probably.
\end{quote}
Carrie and Mara both noted that the XAI overlay was manageable assuming that the user was willing to invest the necessary effort:
\begin{quote}
	\textbf{Carrie:} I have got better at it. At first I wasn't quite understanding it, but it is not difficult to understand. But it does take a little bit of cumulative [knowledge]. Just getting used to it.
\end{quote}
\quotesep
\begin{quote}
	\textbf{Mara:} However, as an intelligent human being I can be taught to read one, even if it's not the best graph ever. Do you know what I mean? If you keep showing me every month or every whenever it is when I look at this regularly, a graph, then I know immediately how to read it.
\end{quote}
Barış and Ender felt that the detailed information available in the XAI overlay detracted from the primary benefit of using an AI feature, which for them was convenience:
\begin{quote}
	\textbf{Barış:} It is not for me because I like the way AI makes life easier, when I have got four graphs for a radiator it is a bit too much for me.
\end{quote}
\quotesep
\begin{quote}
	\textbf{Ender:} Generally people just want the results, so if I am using some AI based or smart thing I want to see the results of what they did. So, for example, for the robot cleaner everybody is happy with that because it cleans a lot, when you just open the trash bin it is full so you become happy, ``This cleans my house,'' so I don't need the graphs.
\end{quote}
Kevin and Isaad similarly reported a lack of curiosity in how the AI feature worked:
\begin{quote}
	\textbf{Kevin:} As long as I feel comfortable, I won't. I don't think I want to know, as long as I feel comfortable. Because I make changes three times a day, so I think it's quite frequently, so I don't think I will just check how the algorithm will work.
\end{quote}
\quotesep
\begin{quote}
	\textbf{Isaad:} My experience is humans don't interact like that with it. They make a decision on, ``Do I trust it, or do I not trust it?'' That decision is made very quickly, very early before they even start interacting with anything. Everything else after that, you're not thinking about the AI, you're thinking about, ``What does this change for me and what do I do?''
\end{quote}

Several users suggested ways to improve or replace the XAI overlay.
Maya was most favourable but felt that the first chart would be clearer if datapoints included labels to indicate when the corresponding inputs occurred:
\begin{quote}
	\textbf{Maya:} I would want to see the time, maybe little time tags on top of each dot.
\end{quote}
Mara and Sam both felt that an embedded video tutorial would offer users an easier (re)introduction to the XAI overlay.
In contrast, Theo suggested that two levels of granularity would help to avoid information-overload:
\begin{quote}
	\textbf{Theo:} Yeah, so maybe two levels of detail just broadly what does it do and then if you want to find out more into the specifics like this level, you could go into it.
\end{quote}
Ender felt that the XAI overlay should be much simpler:
\begin{quote}
	\textbf{Ender:} You need to have some background to understand what these things are so it could be much more simpler in my opinion.
\end{quote}
Açelya felt that textual information or simple graphics would be more accessible for lay users compared to charts:
\begin{quote}
	\textbf{Açelya:} So maybe just small tips as text based because normal people don't want to spend time to understand the graphics and the numbers.
	Since [Ender] is an analytical person I think he will have some time to understand but normally average people don't have time and maybe don't have the background. So maybe some text tips and text based or just basic visuals can be much more in control.
\end{quote}
Several of these suggestions were addressed (unsuccessfully) within the design of Squid and the study itself.
For example, the suggestion by Theo was reflected in the charts on the profiles page and timeslot overlay, which had more detailed variants as the third and fourth chart on the XAI overlay.
No user made a connection between charts on the XAI overlay and charts elsewhere.
Likewise, the suggestion by Açelya was reflected in the gauge (graphic) and notifications page (text-based summaries).
Finally, the suggestion by Mara and Sam was reflected in the four videos sent to participants during Weeks 1--2; although they were not embedded in the interface, one focused entirely on the XAI overlay.
Several users reported that they did not watch these videos.

\subsection{Diagnosing AI Attacks (Phase 2)}\label{sec:diagnostics}
At the conclusion of Phase~2, during Interview 3, participants were asked to reflect on their most-used features in Phase~2 and to compare those with their previous rankings for Phase~1.
Participants were not asked specifically to update their rankings; although some chose to do so, others focused only on their top two or three most-used features.
Since Phase~2 was focused on AI attacks, the most-used features during this phase can be seen as an indication of what features were used as diagnostic tools.
Both the notifications page and profiles page were reported by all users as among their more-used features during Phase~2, with the notifications page in particular being frequently reported as the most-used feature.
In the previous rankings for Phase~1, the notifications page was second lowest and the profiles page was third highest; both features saw an increase in significance during Phase~2, with the notifications page in particular seeing a significant increase.
On the opposite end, the timeslot overlay was often highlighted as a feature that was used less frequently compared to Phase~1, and in several instances was reported as the least-used feature in Phase~2.
The quick access panel was often overlooked by users when discussing Phase~2, likely due to its prominence in the interface and their preoccupation with AI attacks; when prompted, however, users tended to report that it was still frequently used.
The schedule page was also largely overlooked by users; when prompted, they tended to report that its use had remained steady, although during Phase~1 it was only moderately used and we know from user activity data (see Table~\ref{tbl:user_activity}) that no schedule-edits ocurred during Phase~2.
Use of the XAI overlay tended to remain low in both phases.
To summarise, the features from most-used to least-used during Phase~2 were as follows: (6) quick access panel, (5) notifications page, (4) profiles page, (3) schedule page including edit overlay, (2) timeslot overlay, and (1) XAI overlay.
This includes a significant increase in use of the notifications page from Phase~1 to Phase~2, a moderate increase for the profiles page, a moderate decrease for the schedule page including edit overlay, a significant decrease for the timeslot overlay, and no change for the XAI overlay.
These observations are consistent with how participants typically described their strategies for diagnosing and mitigating suspected AI attacks.

\subsubsection*{Diagnostic Strategies}
As outlined in Section~\ref{sec:attacks}, there were various direct and indirect indicators of AI attacks within the web interface, including the quick access panel, timeslot overlay, profiles page, XAI overlay, and notifications page.
Theo and Naadir both used the quick access panel as a convenient initial check to identify anomalous setpoints:
\begin{quote}
	\textbf{Theo:} It's the easiest way to find out what the temperature in the room is.
	That's, sort of, my starting point every time.
\end{quote}
\quotesep
\begin{quote}
	\textbf{Naadir:} And that one always just to see at the end of the day, we've still got the temperature on the right setting.
	I guess that was always where I'd go to and, sort of, look and see what was going on.
\end{quote}
Theo adopted a strategy that was reactive rather than proactive: if he noticed abnormalities on the quick access panel, such as an unusually low setpoint, he would then go to the notifications page to investigate further.
Mara used the notifications page, followed by the gauge, followed by the quick access panel; unusually the quick access panel was her final step.
Simone used the chart on the profiles page followed by the notifications page, and her strategy appears to have been proactive rather than reactive:
\begin{quote}
	\textbf{Simone:} So I just clicked on each of the profiles, weekends, nights, and I just checked that the line was going the right direction, it was going down.
	Then, subsequently to that, I would go into the notifications and kind of check, scroll down back to the last time I remembered I had looked checking there wasn't anything.
\end{quote}
Theo described the notifications page as ``straightforward'', while Simone said it was useful because it allowed her to ``see back in time''.
Kevin felt that inspecting the notifications page was the most reliable and convenient strategy: 
\begin{quote}
	\textbf{Kevin:} Because this one, this notification, I can check the target temperature and I can check the price at the same time.
	That normally will check this notification, if I got--- if I feel anything weird or strange.
\end{quote}
Simone used the notifications page to look for entries containing information on previously adopted setpoints as well as the context in which they were adopted (e.g.\ time, price).
Isaad relied entirely on the notifications page and limited his focus to detecting anomalous user-category events:
\begin{quote}
	\textbf{Isaad:} Yes, so the notification, because it's a log, so if there's something on that log that you haven't actually--- you know it's an attack.
\end{quote}
This strategy would be sufficient for A-1 attacks but would not work for A-2 or \mbox{A-3} attacks.
Wiola used the notifications page followed by the gauge.
Similar to Isaad, Wiola sought to detect anomalous user-category events, but she appears to have adopted a more permissive strategy than Isaad (looking for ``anything unusual''):
\begin{quote}
	\textbf{Wiola:} So I felt--- or just to see if there were user entries that wasn't us.
	So that was typically my go-to if I wanted to see if there was anything unusual or anything unusual had happened over the past 24 hours, I went to that one first \dots
\end{quote}
Carrie reported that she inspected system-category events on the notifications page but did not attribute this strategy to successfully detecting any attacks:
\begin{quote}
	\textbf{Carrie:} Even though that was most detailed and you had to read it carefully, that was the one where you could actually really see what had happened. It was interesting that it was half-hourly because, even if things hadn't really changed, it was still interesting to see the notification.
\end{quote}
Maya placed significant emphasis on the notifications page:
\begin{quote}
	\textbf{Maya:} I always checked the notifications.
	If I didn't see anything on the notifications, I wouldn't consider it as an attack.
\end{quote}
While the notifications page was sufficient on its own to detect all three attacks, it was not the only source of attack indicators.
Sam also placed significant emphasis on the notifications page but his reason was that he struggled to understand other features of the web interface, including the profiles page and XAI overlay:
\begin{quote}
	\textbf{Sam:} I don't look at the graphs, the charts, because unfortunately I didn't really understand that much of it.
\end{quote}

Wiola used a proactive rather than reactive strategy by inspecting different profiles on the profile page to (retrospectively) identify anomalous gauge settings:
\begin{quote}
	\textbf{Wiola:} I thought it was interesting to, sort of, click through the different profiles
	\dots
	But then the profile was--- the dial wasn't in the right place.
\end{quote}
\quotesep
\begin{quote}
	\textbf{Wiola:} So the profile was obviously off, but we hadn't noticed when it	had happened because we weren't in the house.
\end{quote}
Ernesto focused on the gauge and did not pay much attention to the quick access panel:
\begin{quote}
	\textbf{Ernesto:} To detect, yes, and also, because in our case, me, I was always checking the data, because checking the temperature, as I said, I considered that parameter not something I create for us, 
	\dots
	it was always the gauge, what was drawing me, which is great.
\end{quote}
Barış felt that the gauge was the only way to detect A-2 attacks:
\begin{quote}
	\textbf{Barış:} Actually you can catch, with just the gauge, without being there.
	This is the only way I think that we can catch.
\end{quote}
In contrast, Isaad, Theo, and Maya all reported that they did not use the gauge.
Maya said that she ``didn't like it very much'', while Theo found it difficult to understand:
\begin{quote}
	\textbf{Theo:} I didn't really look at the price sensitivity graphic very much, and it's something I'd never done in the first period either.
	It was a part of it that I found a little bit, I don't know, complicated or not easy to take onboard and digest.
\end{quote}
\quotesep
\begin{quote}
	\textbf{Theo:} It's almost like you had to understand how the whole thing worked to be able to understand how the graphic was, rather than the graphic showing you how it worked, if you understand what I mean.
\end{quote}

Carrie was the only user who referenced the XAI overlay in relation to diagnosing AI attacks; she echoed her previous comments from Phase~1 in that she struggled to understand the confidence regions.
Only Maya reported using the XAI overlay more during Phase~2 than Phase~1:
\begin{quote}
	\textbf{Maya:} It helped me understand what's happening. Yes, I made sense of it I guess.
\end{quote}
However, Maya did not attribute this to the objective of diagnosing AI attacks specifically, and Barış implied that this was not the case:
\begin{quote}
	\textbf{Barış:} Maybe we missed one of the attacks because of not checking this.
\end{quote}
Sam and Barış both implied that they did not use the XAI overlay because the information was available elsewhere (and presumably in a format they felt was more accessible):
\begin{quote}
	\textbf{Sam:} But also, most of my information---correct me if I'm wrong---I can get from the logs.
\end{quote}
\quotesep
\begin{quote}
	\textbf{Barış:} Because I don't need to have a look for this screen to catch the attack or to understand the attack.
	I don't need to have a look for that page.
	I'm choosing the shortest way.
\end{quote}

\subsubsection*{Unreliable Stategies}
There is strong evidence that some users adopted unreliable strategies for diagnosing AI attacks by rigidly interpreting atypical system behaviour as attack indicators.
For example, Simone and Ernesto both appear to have adopted unreliable strategies for interpreting the chart on the profiles page, e.g.\ where attacks would only be indicated by a zero or negative price sensitivity:
\begin{quote}
	\textbf{Simone:} The top two, poisoning, at least I knew, kind of for sure, because the [line] would be going up or straight, it wouldn't go down.
\end{quote}
\quotesep
\begin{quote}
	\textbf{Ernesto:} Yes, that when slope is so obvious, it means they have altered the data, or things, because you expect it to be more steady, very low slope. When you see a much higher one, it is quite clear, anyway we are not--- because that one, was not like the initial ones where kind of were an obvious thing.
\end{quote}
Klara appears to have adopted an unreliable strategy for interpreting the gauge, e.g.\ in one instance concluding that an attack had occurred merely because the dial pointed to a red segment:
\begin{quote}
	\textbf{Klara:} Then when we checked it, the profile, the temperature, everything was fine, until actually we checked the---, it was on the red, so it was the only indication that actually something went wrong.
\end{quote}
On the other hand, Carrie and Maya both recognised that the red segments of the gauge were not necessarily reliable indicators of an attack:
\begin{quote}
	\textbf{Carrie:} The gauge was and wasn't useful because I think, also, I was setting the temperature.
	It was saying I was very sensitive to price changes so it was quite often across in the red and that's partly---
	I guess, if you were using the system in reality, you'd set it to hit the temperature you would want and I thought---
	I think that's where I found things a bit difficult to spot because it was often in the red.
\end{quote}
\quotesep
\begin{quote}
	\textbf{Maya:} But the thing is we could, if we mess with our temperature too often, also mislead that gauge thing so I would never be sure if it's off because of me or because of an attack, because I could miss---
\end{quote}
Only Ender reported to have seen the \emph{unknown} setting of the gauge (see Section~\ref{sec:software}) and, similar to Klara's interpretation of the red segments, concluded that this indicated an attack because it was ``not normal''.

\subsubsection*{Detections \& Misdetections}
Ernesto first detected an A-1 attack while inspecting the profiles page, subsequently confirming his suspicion by checking the notifications page:
\begin{quote}
	\textbf{Ernesto:} Then I noticed that as soon as I did that one of the profiles changed.
	Then, I went to the notifications and then the user said, ``You have done this.''
	I'm like, ``Ah, I haven't done that.'' I just kind of catch it on the spot, I believe.
\end{quote}
Carrie detected an A-1 attack by identifying anomalous user-category events on the notifications page:
\begin{quote}
	\textbf{Carrie:} I was actually looking for where the temperature had been set because, a couple of times, it set it to seven degrees and that
	was really obviously what was going on.
\end{quote}
Comments from Naadir suggest that he was unaware of the possibility of indirect indicators of attacks, focusing instead on environmental indicators for A-2 attacks:
\begin{quote}
	\textbf{Naadir:} Not only that, I think realising one of the attacks was one that didn't record in the notifications.
	That, sort of, maybe changed my look at it in a way.
	It was more really listening to that as opposed to checking them, the notes and the notification.
\end{quote}
Debora detected an A-2 attack by observing anomalous behaviour on the gauge:
\begin{quote}
	\textbf{Debora:} There was, I think, one that I noticed that the gauge was not correct but I didn't see anything on the notification.
\end{quote}
Conversely, Mara noticed anomalous gauge behaviour but her comments suggest that she did not feel this was a reliable attack indicator:
\begin{quote}
	\textbf{Mara:} Only once, I saw the gauge going haywire.
	I don't think the other times something was happening.
\end{quote}
Explaining how they successfully detected A-3 attacks, Simone and Maya both felt that it was due to their understanding of the AI model itself, which was supported by the chart on the profiles page:
\begin{quote}
	\textbf{Simone:} Whereas the third one, it was logical.
	If you looked at the gauge and you looked at the [line], it was going the right	direction.
	Because the price was so high, it was being set very low.
	So this kind of made me doubt, ``Is this right?''
	Then I thought, ``Ah, this is another way to do an attack, they're going to change the price. That's what I thought.
\end{quote}
\quotesep
\begin{quote}
	\textbf{Maya:} So I checked the overall temperature/price ratio and I thought, ``Maybe it's thinking the price is really high so it had to make it
	like 7°.''
	So I was thinking it's the logic of the AI, basically.
\end{quote}
Maya also avoided a false positive by inspecting the chart on the profiles page:
\begin{quote}
	\textbf{Maya:} I'd seen the temperature was like 15° and we don't really arrange it to 15° so I was suspicious if this is an attack. But then I checked the
	graph, I've seen that if the price is about, let's say, 30p the temperature would be 15°. So the graph showed it to me, so then I understood, ``Oh, this is not an attack, this is accurate in terms of the graph.''
\end{quote}
Theo failed to detect any A-3 attacks and felt that his adopted strategy (of checking the quick access panel followed by the notifications page) would have only worked for A-1 attacks.
However, Simone recognised that Theo's strategy would have worked for A-3 attacks as long as the web interface was inspected while the attack was still ongoing:
\begin{quote}
	\textbf{Simone:} I think it's more a timing thing because you could have been in the living room and noticed it was really cold,
	opened the tablet, and it would have said, ``The price is 7° because it's £100 per kilowatt.''
	So your way of diagnosing would have still worked.
	It's just the timing that when you looked it didn't happen to be this type of attack.
\end{quote}
Discussing why they failed to detect any A-3 attacks, Wiola and Naadir both recognised that the profiles page would not necessary help, while Naadir recognised that the quick access panel was a suitable alternative:
\begin{quote}
	\textbf{Wiola:} Well, on the gauge, would have not been effective?

	\textbf{Naadir:} No. It [would not have] been on there, right?
	It would've been on that front page where it says current prices.
\end{quote}
Mara failed to detect A-3 attacks because they did not conform to her adopted strategy, while Debora suspected but ultimately overlooked an A-3 attack:
\begin{quote}
	\textbf{Mara:} I don't think I expected anything, I was just imagining that---
	I guess I detected what I expected, I don't think this was part of---
	That was quite different. I knew there was something to do with prices that was going wrong, which I was telling you about, and how the ratio was changing, without the temperature changing.
	I didn't realise the notifications tool was attacked.
	That should have been a guess, but I didn't think of it.
\end{quote}
\quotesep
\begin{quote}
	\textbf{Debora:} To be honest, once I checked the notification, I did notice
	some weird thing with the rise in temperature but yes, I didn't
	think it was an attack because I didn't notice anything else.
\end{quote}

\section{Discussion}\label{sec:discussion}

\subsubsection*{Sample size}
Originally we aimed to recruit 20 households from the social housing sector in England.
This proved unachievable, both in raw numbers and in recruiting from that sector specifically.
Our final sample consisted of 18 participants across 10 households, with three households from the private rental sector and the rest being owner-occupiers.
While this sample size ($n = 18$) was below our target, it nonetheless exceeds the median sample size ($n = 11$) discussed in Section~\ref{sec:related_work} for previous qualitative XAI user studies from the literature.
Even counting participants from the same household as a single unit ($n = 10$), our sample size remains similar to those previous studies.
Compared to the SmartThermo study~\cite{Alan:CHI:2016,Shann:AAMAS:2017}, which had $n = 30$ participants based on one participant per household, our sample size is lower.
However, compared to the number of participants using the indirect learning design in that study ($n = 10$), which is the design replicated in Squid, our sample size exceeds that based on participants ($n = 18$) and remains similar based on households ($n = 10$).
Saturation analysis was not conducted a priori due to various logistical constraints, including the longitudinal nature of the study and the need to conclude before the start of warmer weather when home heating would not be in use (e.g.\ by April 2023).
However, as shown in Section~\ref{sec:results}, analysis of interviews a posteriori suggests that a level of saturation was achieved.

\subsubsection*{Simulated prices}
The AI method used by Squid and SmartThermo assumes that a user's preferred setpoint depends on the current energy price.
However, the respective field studies for these applications both relied on prices that were \emph{simulated}.
In the case of Squid, prices were taken directly from historical data for the Agile tariff by Octopus Energy.
In the case of SmartThermo, which was evaluated prior to the release of the Agile tariff, prices were instead derived from historical electricity spot prices~\cite{Alan:CHI:2016,Shann:AAMAS:2017}.
Regardless the source of simulated prices, what they mean for each study is that users were shown prices within the applications that did not reflect what those users were actually paying for energy.
In essence both studies were thus asking users to immerse themselves in a hypothetical scenario: to behave during the study as if the simulated prices were real prices.
From our interviews there is evidence that some users struggled with this request:
\begin{quote}
	\textbf{Klara:} To be honest, I completely forgot actually about the price, I was actually so focused on the temperature.
	Now, I recall maybe talking about it.
	Like, ``Yes, there are prices, but they are not real, but this is a thing that you actually have to know what--- you have to be aware of the prices \dots''
\end{quote}
\quotesep
\begin{quote}
	\textbf{Ernesto:} Knowing that they're fake, you really bring something in your head, that if they were real prices, you probably would look more into detail.
\end{quote}

\begin{figure}[tbp]
	\centering
	\includegraphics[width=0.5\textwidth]{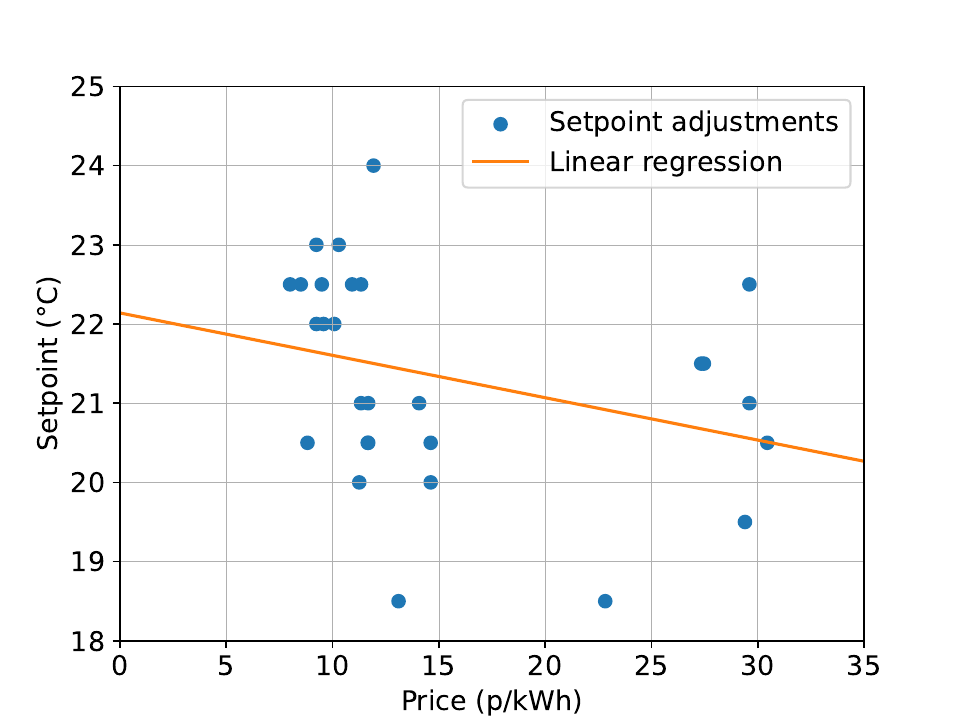}
	\caption{Results for linear regression on setpoints adjustments ($n = 28$) for the user with lowest $p$-value ($p = 0.109$).}
	\label{fig:rational}
\end{figure}

In~\cite{Shann:AAMAS:2017} the authors conducted a linear regression analysis of user setpoint adjustments during their study with respect to simulated prices.
For 20\% of users, they found statistically significant results ($p < 0.05$) that were said to confirm a hypothesis that users would behave in a way that was \emph{economically rational}, i.e.\ for these users there would be a statistically significant negative slope.
For the remaining 80\% of users, results were not statistically significant.
When interpreting these results it is important to remember that users were participating in a scenario and it is unlikely that all users would have fully immersed themselves in that scenario; if those same users were faced with real prices, then we might reasonably expect stronger results, given that the premise is a simple one of economic rationality.
We conducted the same analysis for our own study but found no statistically significant results.
See Figure~\ref{fig:rational} for the inputs and best-fit line for the user with lowest $p$-valve from that analysis.
Our results thus do not confirm the results from~\cite{Shann:AAMAS:2017}.
A possible explanation is that many Squid users did not successfully immerse themselves in the scenario.
However, for those users who appeared to succesfully engage with the simulated prices, most still demonstrated low price sensitivity:
\begin{quote}
	\textbf{Ender:} I am looking at the price but to understand what is the general price in that period of time. So for informative reasons I check it. If I feel cold or things like that I just increase the temperature but I never increase it above 21.5, so I generally set it to 21.5. Most of the days now it is set to that temperature so I think it learned but in some cases it was lower so I was checking, but I didn't just set my usage depending on the price in general.
\end{quote}
\quotesep
\begin{quote}
	\textbf{Simone:} But I do look at it, yes, I do. But I would say, yeah, I guess number one is are the boys here and how cold is it? And then I look at the price and I think, ``Okay, I'm going to increase by 0.5 or---''
\end{quote}
While this outcome may affect the accuracy of learned preferences, it should not in principle affect aspects of the study related to XAI and AI attacks, since all users still demonstrated awareness that the AI feature depended in those prices.

\subsubsection*{Cost estimates}
One of the most-liked features in SmartThermo was its \emph{30-day estimated cost}~\cite{Alan:CHI:2016}.
It is easy to see why cost estimates would be appealing to users.
Acording to~\cite{Shann:AAMAS:2017} the aim in SmartThermo ``was not to provide the most accurate 30-day cost prediction possible, but to create enough realism such that the users could immerse themselves''.
Cost estimates were based on the user's learned setpoint schedule combined with a ``relatively simple'' thermal model~\cite{Rogers:BuildSys:2013} that accounted for building thermal dynamics and weather forecasts.
We investigated the inclusion of similar cost estimates in Squid but unfortunately our experimental results proved both unstable and inaccurate.
On the one hand, unstable cost estimates have usability implications, since anomalous fluctuations may confuse users.
On the other hand, inaccurate cost estimates have ethical implications, as evidenced by the following quote from a user of SmartThermo~\cite{Alan:CHI:2016}:
\begingroup
\addtolength\leftmargini{-0.065in}
\begin{quote}
	I've watched also my estimated cost each day, to see whether it varied at all.
	I had taken an interest in it, every day really, I've become almost fixated by it.
\end{quote}
\endgroup
\noindent This quote highlights the degree to which cost estimates may influence user behaviour, regardless of whether those cost estimates are accurate; the developers of SmartThermo made no particular claims about the accuracy of its cost estimates yet this user seems to have become heavily invested in them. 
However, if cost estimates are inaccurate, then in the worst case users might adopt behaviour they believe to be cost-saving that in reality serves to increase their true costs.
Estimating costs for our field study was especially challenging because true costs were associated with a TRV (i.e.\ a single radiator) rather than a central thermostat (i.e.\ the entire dwelling).
For these reasons we ultimately decided to omit cost estimates from Squid.
Nonetheless, the lack of cost estimates was an issue raised by some users and may have impacted user engagement:
\begin{quote}
	\textbf{Klara:} I think what I would expect actually, from the technology is to tell me, on a weekly basis, the way that you have been using it, that this is what you can expect on an average monthly basis. The cost.
\end{quote}
\quotesep
\begin{quote}
	\textbf{Ender:} Also we didn't understand what will be the total benefit, so the price is something but I don't know if I just used the AI to control things what will it [save], £1 per month, £10, £25, so it depends. If it is £1 or £2 I don't care.
\end{quote}
To the best of our knowledge, no consumer smart thermostat includes cost estimates, which seems to corroborate our experience that accurate cost estimation in this setting remains challenging and error-prone.

\subsubsection*{Profile resets}
The ability to reset profiles was built into the design Squid so as to allow users to retain a degree of control over its AI feature.
Profile resets also served as a primary mitigation against suspected A-1 and A-2 attacks and was presented as such during user training.
In Section~\ref{sec:software} we noted that a similar feature was included in SmartThermo but results showed that it tended to be over-used, with the effect being to limit preference learning during the study.
We then suggested that introducing temporality into the AI feature would reduce this over-reliance on profile resets.
Table~\ref{tbl:user_activity} shows that during Phase~1 of our study (i.e.\ prior to the commencement of AI attacks) users made over 28 setpoint adjustments on average before opting to reset a profile.
This result compares favourably to SmartThermo, where on average users of its direct and indirect learning designs made, respectively, less than 3 and 13 setpoint adjustments before opting to reset~\cite{Shann:AAMAS:2017}.
Our results are thus consistent with the hypothesis that introducing temporality would have a positive effect on the over-use of profile resets seen in SmartThermo.

As noted in Section~\ref{sec:results}, the number of profile resets increased sharply between Phases 1 and 2, which we said was unsurprising because of the occurrence of AI attacks during the latter phase.
However, Table~\ref{tbl:user_activity} also shows a relative decline in the number of setpoint adjustments between Phases 1 and 2, particularly when factoring in their durations (i.e.\ three and four weeks, respectively).
Interviews suggest that the increase in profile resets during Phase~2, as a mitigation against suspected AI attacks, had a side-effect of causing some users to disengage from the AI feature itself:
\begin{quote}
	\textbf{Ernesto:} I gave up, because what's the point, when you do reset and you end up spending all this effort \dots
	
	\textbf{Interviewer:} \dots Was it because of the effort involved in retraining it or was it because you knew that there might be another attack?

	\textbf{Ernesto:} Both, because you know that it doesn't matter how much---, at some point you get an attack again, you may need to reset---.
\end{quote}
Similar to previous results for SmartThermo, profile resets thus served to limit preference learning in Squid during Phase~2 such that the default model has an inflated role in setpoint prediction.
However, while the result for SmartThermo was likely due to user frustration caused by limitations of the preference learning itself, the result for Squid in Phase~2 was likely due to user frustration caused by the inconvenience of AI attacks.

\subsubsection*{Client device}
The Squid web interface was originally developed for desktops and tablets, with the expectation that participants would access it using personal devices.
Having a suitable personal device would thus have formed part of the selection criteria during recruitment.
It was decided early in development that smartphones would not be supported due to their limited screen real estate.
Throughout development the web interface was tested on various devices, including a desktop monitor (Dell U2719D), a laptop (MacBook Pro), and a standard-sized tablet (iPad Air).
However, a decision was made rather late in development to provide each household with the same tablet (Galaxy Tab).
This was intended to ensure a uniform experience for all users.
Unfortunately several users reported usability and accessibility issues in relation to the tablet: 
\begin{quote}
	\textbf{Debora:} And actually, sometimes it can be a little bit difficult on the tablet to see the exact number [on the setpoint controller].
\end{quote}
\quotesep
\begin{quote}
	\textbf{Carrie:} I didn't find the graphs, in some ways, that helpful because the daily temperature-to-cost one seemed to be almost the same pattern every day.
	I was looking at it, also, on the tablet and it was actually quite small so you couldn't really perceive big variations in it so, actually, it was the notifications that were more useful.
\end{quote}

The selected tablet proved to offer significantly less screen real estate than had been anticipated during prior development.
Table~\ref{tbl:clients} compares the screen real estate available on the tablet used by participants with the devices previously used during development.
For example, the iPad Air has over 1.5 times the virtual screen real estate compared to the Galaxy Tab, and almost double the physical real estate.
Compared to any device that had been tested previously, the Galaxy Tab was able to display less information on-screen at any one time.
Compared to the iPad Air, the same information on-screen would also appear smaller.
Aspect ratios of 1.68:1 for the Galaxy Tab and 1.33:1 for the iPad Air meant that vertical real estate on the Galaxy Tab was much narrower in landscape mode than on the iPad Air.
At the same time, the Galaxy Tab offered more screen real estate than a standard smartphone, so it was still technically within the parameters of the study.
For example, the Galaxy Tab has over double the physical screen real estate compared to an iPhone SE, and over 3.5 times the virtual real estate.
As such, the Galaxy Tab has characteristics placing it somewhere between a smartphone and a standard-size tablet.

\begin{table}[tbp]
	\centering
    \scriptsize
	\caption{Screen real estate for client used by participants vs.\ earlier in development.}
	\label{tbl:clients}
	\begin{tabular}{l r r r r r r}
		\toprule
		\multirow{2}{*}{\textbf{Device}} & \multirow{2}{*}{\textbf{Aspect ratio}} & \multicolumn{2}{l}{\textbf{Diagonal length}} & \multicolumn{2}{l}{\textbf{Area vs.\ Galaxy Tab}} \\
		& & Physical (inch) & Virtual (pixels)$^{\dagger}$ & Physical & Virtual$^{\dagger}$  \\
		\midrule
		Galaxy Tab & 1.68:1 & 8.7 & 1040.43 & 1:1 & 1:1 \\
		\midrule
		iPad Air & 1.33:1 & 10.5 & 1390 & 1.95:1 & 1.59:1 \\
		MacBook Pro & 1.6:1 & 13.3 & 1698.12 & 2.72:1 & 2.39:1 \\
		Dell U2719D & 1.78:1 & 27 & 2937.21 & 7.74:1 & 9.35:1 \\
		\bottomrule
	\end{tabular}\\
	$^{\dagger}$Based on scaled resolution rather than native resolution.
\end{table}

Particularly affected on the tablet was the XAI overlay, where it was no longer possible to display all four charts on-screen at the same time.
As discussed in Section~\ref{sec:xai}, this particular feature allowed users to observe the impact of their inputs on model parameters and model predictions simultaneously.
The narrow aspect ratio of the Galaxy Tab was exacerbated by fixed menu bars enforced by the platform, further reducing already-limited vertical real estate in landscape mode.
Attempts were made to tailor the web interface specifically to the Galaxy Tab.
For example, the schedule page and XAI overlay were made scrollable, and the navigation menu at the bottom of the interface was set to be always-visible.
However, in spite of these changes, interviews suggest that limitations of the Galaxy Tab still caused problems for some users (especially Carrie) during the study, although the actual functionality was the same regardless of client device.
It is likely that these issues impacted perceptions of affected components, including the XAI overlay and charts shown elsewhere, while increasing the significance of other components, such as the notifications page.
It was acknowledged by users that these issues were specific to small devices such as the Galaxy Tab:
\begin{quote}
	\textbf{Isaad:} I mean, on the laptop you've got both of them [quick access panel and schedule page] at the same time, which is better.
\end{quote}
On the positive side, some users found that having a dedicated device to access the web interface served as a useful reminder to engage with Squid throughout the study:
\begin{quote}
	\textbf{Simone:} Whereas, there, it's like this special tool which is specific for that, I found that it was a good reminder. I guess the mobile phone is a reminder for so many things, I don't know if I would have remembered so well.
\end{quote}
Thus, while adopting the Galaxy Tab proved technically challenging and likely increased negative perceptions of some web interface components, it had the benefit of ensuring a uniform experience for users and may have encouraged user engagement.

\subsubsection*{Hardware setup}
SmartThermo was designed for a central thermostat whereas Squid was designed for a TRV.
As mentioned in Section~\ref{sec:hardware} however, Squid could be easily adapted to work with a central thermostat.
For the most part, this would be achieved simply by replacing the chai-data-sources software component mentioned in Section~\ref{sec:software}.
The decision to use a TRV rather than a central thermostat was made for pragmatic reasons rather than technical ones.
Firstly, installing a central thermostat is more involved than installing a TRV, since the former must be wired to a boiler while the latter is simply screwed on to a radiator.
Our expectation was that a TRV would lower the barrier to entry and thus ease recuitment, due to its less invasive installation.
Secondly, a central thermostat affects the entire dwelling whereas a TRV only affects a localised area (e.g.\ a single room).
Given that our study involved AI attacks, we concluded that a TRV represented a more acceptable level of risk since it would limit any potentially harmful effects on users following the attacks.
However, interviews suggest that this choice did have some negative impact on user perceptions of the significance and usefulness of Squid:
\begin{quote}
	\textbf{Theo:} Yeah, it's quite difficult getting your head around the different layers because obviously if the boiler's not on the whole system's just off.
\end{quote}
\quotesep
\begin{quote}
	\textbf{Maya:} I think for me it is only one radiator so it couldn't harm that much, I think that was one thing.
	I think if it was installed in every radiator at home I would be more onto it, I couldn't have trusted that easily probably.
	Because it is just one thing it is fine if something is going wrong, it wouldn't affect me that much.
\end{quote}
Looking at consumer examples, it is unusual to find a smart TRV that is not coupled with a smart (central) thermostat, e.g.\ the Netatmo Additional Smart Radiator Valve assumes that the Netatmo Smart Thermostat is present, as noted in Section~\ref{sec:hardware}.
However, there is no technical reason why a smart TRV cannot operator on its own, e.g.\ traditional TRVs are agnostic to the presence of a central thermostat.
Some of the confusion expressed by users during interviews perhaps explains the lack of consumer examples of this setup.
In terms of the impact this setup had on the study, there is certainly interview evidence to suggest that it impacted user behaviour, such as to give users less incentive to address inadequate learned preferences.
Conversely, there is no particular evidence to suggest that the setup impacted user engagement with the study itself.

\subsubsection*{AI attacks}
The AI attacks simulated during the study were based on an existing taxonomy of attacks on machine learning~\cite{Pitropakis:CSR:2019}.
In all cases the simulated attacks were designed to impact users via the setpoint learning feature, specifically to cause the system to adopt a setpoint lower than it would otherwise have adopted.
Interviews suggest that users largely did not find these attacks plausible due to the perceived effort required from the attacker versus the potential rewards available to the attacker:
\begin{quote}
	\textbf{Carrie:} I find it quite hard to imagine why people would be attacking individuals apart from, somehow, to get into the banking systems but it seems like a very complex thing to have to do for a relatively small financial gain.
\end{quote}
\quotesep
\begin{quote}
	\textbf{Sam:} Yes it is, because there is only a maximum you could ever go on one radiator, so it's never going to go to super unbearably hot that I feel like I have to wake up in the middle of the night.
	Whereas a breach of data for any computer, and when I say computer I mean any device, Alexa, mobile or whatever.
	That is a bit more sensitive, because we are talking about security.
\end{quote}
Participants generally framed the AI attacks on Squid in terms of harrassment of the user, but felt that the most plausible attacks would be those offering the attacker with a clear source of financial gain.
When participants were prompted with the possibility that the AI attacks on Squid could form part of a large-scale attack on the energy grid, they generally felt that this was more plausible due to the significance of its impact.
However, in this instance participants generally felt that the software owner and/or energy provider should bear the responsibility of defending against the attacks, rather than end-users.
Interviews provide no particular evidence that the plausibility of AI attacks impacted user engagement during the study.
Nonetheless, the responses raise interesting questions for future work in terms of when and how best to protect lay users from the negative effects of AI.

\section{Conclusion}\label{sec:conclusion}
This paper strengthens existing XAI research by contributing a major in-the-wild qualitative study on XAI for lay users.
As illustrated in Section~\ref{sec:related_work}, few comparable studies have been reported in the literature previously.
The study focused specifically on XAI for lay users in the context of AI cyberattacks; to the best of our knowledge this represents a novel use case that has not been considered by previous work.
The results offer both positive and negative insights on the usefulness of XAI from a lay user perspective.
Squid users demonstrated improved mental models of the AI feature compared to previous work.
In several instances, consequential decision-making was directly attributable to certain XAI-related features, such as to avoid false positives in diagnosing AI attacks by correctly interpreting certain charts.
Active engagement with XAI-related features remained low throughout the study, especially with the XAI overlay.
This was true even when users reported positive opinions of those features and demonstrated reasonable comprehension of what they were intended to convey.
All charts were generally well-understood with the exception of the second chart on the XAI overlay.
The confidence regions on the XAI overlay were a common source of confusion.
The gauge was poorly understood compared to the slope on the corresponding charts, and was often misinterpreted when diagnosing AI attacks.
The notifications page was most commonly used to diagnose AI attacks but its use was often limited to identifying anomalous user-category events without considering message content.
At the conclusion of the study, Squid users continued to report a desire to understand the AI systems that affect them:
\begin{quote}
	\textbf{Simone:} I'd like to understand in a general manner, I don't know to what extent I will understand it but I
	would like to have a general idea of how it works rather than it being a black box.
\end{quote}
These findings offer important lessons for future XAI research.
The XAI community may overestimate the degree to which lay users may be willing and/or able to engage with existing XAI tools and concepts.
This corroborates recent reflections on the current state of XAI~\cite{Miller:FAccT:2023}.
The importance of traditional user experience (UX) design to XAI should not underestimated; even for AI systems based on simple glass-box machine learnine models, effective UX design remains crucial.

\section*{Acknowledgments}
This work received funding from the EPSRC CHAI project under grants EP/T026707/1, EP/T026812/1, and EP/T026596/1.
The authors wish to acknowledge: 
Rytis Venslovas for developing the Squid web interface; 
Laura Benton for feedback on the design of Squid and input on the home training; 
Ceylan Besevli, Ana Serta, and Sarah Turner for contributing to the study design and data collection; 
and Etienne Roesch for supporting the home installations.

\bibliographystyle{unsrt}
\bibliography{ijis2024}

\section*{Appendix}

\subsection*{XAI Overlay}
There are four tooltips, one for each chart, as follows:
\begin{itemize}
	\item This chart visualises your profile \textbf{inputs} over time, since your last profile reset. Each input is comprised of a \textbf{target temperature} change and the \textbf{energy price} when the change was made. Each input serves to \textbf{update} your AI model.
	\item This chart visualises your \textbf{AI model} over time. The \textbf{best guess} is a learned estimation of your \textbf{preferred temperature (if energy were free)} and your \textbf{price sensitivity}. The \textbf{confidence region} represents uncertainty over the best guess: a larger confidence region means more uncertainty. The AI model is used to make \textbf{predictions} about your ideal target temperature relative to energy price.
	\item This chart visualises your AI model \textbf{predictions} over time. The \textbf{best guess} is a learned estimation of your ideal target temperature relative to energy price. The \textbf{confidence region} represents uncertainty over the best guess: a larger confidence region means more uncertainty. The predictions are used in \textbf{auto mode} to choose your target temperature relative to the current energy price.
	\item This chart visualises the \textbf{energy price shedule} for a given day along with your \textbf{target temperatures} in auto mode for that schedule and this current profile. In reality your target temperatures in auto mode will depend on both the energy price schedule and your \textbf{profile schedule}: each profile has its own AI model with its own predictions, even if energy prices remain the same.
\end{itemize}

\subsection*{Notifications Page}

\subsubsection*{User Category}
There are four notification types within the \emph{user} category as follows:
\begin{itemize}
	\item You set the target temperature to $Y\degree$C ($M$ mode is now active).
	\item You switched to $M$ mode.
	\item You reset profile $P$.
	\item You edited the schedule.
\end{itemize}
Parameters include setpoint $Y \in \{ 7, 7.5, \dots, 30 \}$, mode $M \in \{ \text{on}, \text{off}, \text{auto}, \text{override} \}$, and profile $P \in \{ \text{Nights}, \text{Mornings}, \text{Weekdays},$ $\text{Evenings}, \text{Weekends} \}$.

\subsubsection*{System Category}
There are two notification types within the \emph{system} category as follows:
\begin{itemize}
	\item The system set the target temperature to $Y\degree$C because the current price is $X$p/kWh and the active profile is $P$ where the AI believes your price sensitivity is $S$ and your preferred temperature (if energy were free) is $T\degree$C.
	\item Profile $P$ has been updated because you set the target temperature to $Y\degree$C when the price was $X$p/kWh where the AI now believes your price sensitivity is $S$ and your preferred temperature (if energy were free) is $T\degree$C.
\end{itemize}
Parameters include setpoint $Y$ and profile $P$ (as above) as well as price $X \in \mathbb{R}$, price sensitivity label $P \in \{ \text{Negative}, \text{Very low}, \text{Low}, \text{Moderate}, \text{High}, \text{Very high}, \text{Undefined} \}$, and preferred temperature (if energy were free) $T \in \mathbb{R}$.
Note that price sensitivity labels correspond from left-to-right to the six segments of the gauge as described in Section~\ref{sec:software} such that the final label indicates the special case where the six-point scale is undefined.

\subsection*{Flash Messages}
There are seven flash message types as follows:
\begin{itemize}
	\item System is in $M$ mode
	\item Current price is $X$p/kWh
	\item Active profile is $P$
	\item Target temperature is $Y\degree$C
	\item All profiles are reset
	\item $P$ profile is reset
	\item $D$ schedule is updated
\end{itemize}
Parameters include mode $M$, price $X$, profile $P$, and setpoint $Y$ (as above) as well as day $D \in \{ \text{Monday}, \text{Tuesday}, \dots, \text{Sunday} \}$.

\end{document}